# Handling missing data when estimating causal effects with Targeted Maximum Likelihood Estimation

S. Ghazaleh Dashti[1,2], Katherine J. Lee[1,2], Julie A. Simpson[3], Ian R. White[4], John B. Carlin[1,2], Margarita Moreno-Betancur[2,1]

1 Clinical Epidemiology and Biostatistics Unit, Murdoch Children's Research Institute, Melbourne, Victoria, Australia
2 Clinical Epidemiology and Biostatistics Unit, Department of Paediatrics, University of Melbourne, Melbourne, Victoria, Australia
3 Centre for Epidemiology and Biostatistics, Melbourne School of Population and Global Health, University of Melbourne, Melbourne, Victoria, Australia
4 MRC Clinical Trials Unit at UCL, University College London, London, UK

Corresponding author
S Ghazaleh Dashti, DDS, MPH, PhD
Clinical Epidemiology and Biostatistics Unit
Murdoch Children's Research Institute
Royal Children's Hospital, 50 Flemington Rd, Parkville VIC 3052
Email: ghazaleh.dashti@mcri.edu.au

**Acknowledgements**

This article used unit record data from the Victorian Adolescent Health Cohort Study. We thank the families that participated in the VAHCS, the study research team and the Principal Investigator, Professor George Patton. We also thank Anthony Charsley for his contributions to the coding in the preliminary stages of this work.

**Funding**

Ian White was supported by the Medical Research Council Programme MC_UU_00004/07. Katherine Lee was funded by a National Health and Medical Research Council (NHMRC) Career Development Fellowship (ID 1127984), Julie Simpson by a NHMRC Investigator Grant – Leadership (ID 1196068), and Margarita Moreno-Betancur by an NHMRC Investigator Grant (ID 2009572). The work was supported from the National Health and Medical Research Council Project Grant (ID 1166023). The Murdoch Children's Research Institute is supported by the Victorian Government's Operational Infrastructure Support Program.

**Author Contributions**

All authors participated in planning the simulation and case studies, interpretation of the results, and reviewing and revising the manuscript. SGD and MMB conceptualized and designed the study. SGD performed the simulation and case study analyses and led the writing of the manuscript.

**Data availability statement**

Data from the Victorian Adolescent Health Cohort Study (VAHCS) are not publicly available. Those interested in replicating these findings are welcome to contact the corresponding author, or the VAHCS team (https://www.mcri.edu.au/research/projects/2000-stories/information-researchers). Simulation study codes can be made available upon request to the corresponding author.

**Conflicts of Interest**

The authors declare no conflicts of interest.

**Abstract**

Targeted Maximum Likelihood Estimation (TMLE) is increasingly used for doubly robust causal inference, but how missing data should be handled when using TMLE with data-adaptive approaches is unclear. Based on the Victorian Adolescent Health Cohort Study, we conducted a simulation study to evaluate eight missing data methods in this context: complete-case analysis, extended TMLE incorporating outcome-missingness model, missing covariate missing indicator method, five multiple imputation (MI) approaches using parametric or machine-learning models. Six scenarios were considered, varying in exposure/outcome generation models (presence of confounder-confounder interactions) and missingness mechanisms (whether outcome influenced missingness in other variables and presence of interaction/non-linear terms in missingness models). Complete-case analysis and extended TMLE had small biases when outcome did not influence missingness in other variables. Parametric MI without interactions had large bias when exposure/outcome generation models included interactions. Parametric MI including interactions performed best in bias and variance reduction across all settings, except when missingness models included a non-linear term. When choosing a method to handle missing data in the context of TMLE, researchers must consider the missingness mechanism and, for MI, compatibility with the analysis method. In many settings, a parametric MI approach that incorporates interactions and non-linearities is expected to perform well.

## Introduction

A key component of epidemiological research is causal inference from longitudinal studies, where the objective is often to estimate the average causal effect (ACE) of an exposure on an outcome.(1-5) For a binary exposure (X=1 exposed; X=0 unexposed), the ACE can be defined as the difference in the average potential outcome if all were exposed versus unexposed.(1-5) In the absence of missing data, under the assumptions of conditional exchangeability given a vector of measured confounders Z, consistency, and positivity, the ACE is identifiable from observable data by the g-formula $\mathrm{E}[\mathrm{E}(\mathrm{Y}|\mathrm{X}=1,\mathrm{Z})-\mathrm{E}(\mathrm{Y}|\mathrm{X}=0,\mathrm{Z})]$, where Y is the outcome.(6)

Several singly robust approaches, including g-computation and propensity score methods, and doubly robust estimators, including Targeted Maximum Likelihood Estimation (TMLE), are available for ACE estimation in the absence of missing data. Here, we focus on TMLE, which combines models for the outcome and propensity score.(7-9) A detailed description of TMLE is available elsewhere.(7-9) Briefly, the first step is the same as in g-computation, where a model for the expected outcome conditional on exposure and confounders ($\widehat{\mathrm{E}}[\mathrm{Y}|\mathrm{X},\mathrm{Z}]$) is estimated and used to predict outcome for all records under exposure and no exposure. In g-computation these predictions are directly plugged into the g-formula to estimate the ACE. In TMLE, they are updated using information from the propensity score ($\widehat{\mathrm{P}}[\mathrm{X}=1|\mathrm{Z}]$) (the targeting step), before being plugged into the g-formula.(7) The targeting step in TMLE ensures that the estimator is doubly robust, where only one of the two models (outcome or propensity model) needs to be consistently estimated to ensure consistent estimation of the ACE. Unlike with singly robust methods, data-adaptive approaches (e.g. machine learning methods) can be used for the exposure and outcome models in TMLE as long as the Donsker class condition (requiring that outcome and propensity score estimators do not heavily overfit the data) holds.(8-10) Given these desirable properties, interest in the application of TMLE for ACE estimation is growing.

Missing data are ubiquitous in epidemiological studies and can lead to biased estimates and loss of precision if handled inappropriately.(11) Commonly used approaches for handling missing data in studies using TMLE for ACE estimation include multiple imputation (MI; e.g.,(12)), complete-case analysis (CCA; e.g.,(13)), extending TMLE to handle missing outcome data (e.g.,(14)), and the missing covariate missing indicator (MCMI) approach for handling missing confounder data (e.g.,(15)). No study has compared these methods in terms of bias and precision. Also, while a requirement for valid inference with MI is that the imputation model should incorporate all relationships assumed to hold in the analysis method, (11) the optimal implementation of MI when using TMLE with data-adaptive methods for ACE estimation is unknown. Answers to these questions are key to developing guidance for appropriate handling of missing data in the context of the growing use of TMLE in applied epidemiological research.

In the current paper, we sought to address this knowledge gap using a simulation study based on an illustrative example from the Victorian Adolescent Health Cohort Study (VAHCS). Our interest was to compare the performance of readily available missing data methods to inform current practice. We begin by introducing the VAHCS example, then describe methods for handling missing data with TMLE, present the simulation study we conducted to evaluate and compare the performance of these approaches, then illustrate the assessed approaches in the VAHCS example. We conclude with a general discussion.

## Illustrative example

Our example was based on a previous investigation using data from VAHCS, a longitudinal cohort study of 1,943 participants (1,000 females), recruited in 1992-1993 at ages14-15 years.(16) Data were collected during participants' adolescence (waves one-six) and in young adulthood (1998, wave seven). Investigators aimed to estimate the ACE of frequent cannabis use in adolescent females on mental health in young adulthood, measured using the Clinical Interview Schedule-Revised (CIS-R). We revisited this analysis using TMLE with data-adaptive approaches, using the same confounders considered in the original investigation: parental divorce, antisocial behaviour, depression and anxiety, alcohol use, parental education, all measured across waves two to six.(17) **Table 1** shows descriptive statistics for analysis variables, and age at wave two, which is useful as an auxiliary variable in MI (a predictor of missing values but not included in the analysis method (11)). All variables had some degree of missingness, ranging from 0.1% to 30.8%.

## Methods for handling missing data in TMLE

Approaches proposed for handling missing data when estimating the ACE using TMLE in studies like our example are described below.

### *Non-MI approaches*

*Complete-case analysis (CCA):* only records with complete data for target analysis variables are used.(18) This approach generally leads to loss of precision (19) and, depending on missingness mechanism, may inflict bias.(18)

*Extended TMLE in a sample with complete exposure and confounders (Ext-TMLE+CEC)*: records with missing data for Z and X are deleted. From records with complete data on Y, E[Y|X, Z] is estimated. In the targeting step, the predictions of the outcome are updated using information from models for P[X=1|Z] and P[$M_Y$ =0|X,Z], where $M_Y$ is the missingness indicator for the outcome.(20) Updated predictions for the outcome under exposure and no exposure are obtained for all records, regardless of whether they have missing outcome.(20) The model for $M_Y$ can also be fitted using data-adaptive approaches. With missingness only in the outcome, the extended TMLE method is unbiased under an extended exchangeability assumption ($Y^x \coprod M_Y|X, Z$ and $Y^x \coprod X|Z$ for $x = 0,1$, where $Y^x$ is the potential outcome when $X = x$, and $\coprod$ denotes independence).(21)

*Extended TMLE plus missing covariate missing indicator (MCMI) approach (Ext-TMLE+MCMI)*: missing outcome data are handled using the extended TMLE approach, and missing confounder data by including missingness indicators for the incomplete confounders in the confounding adjustment set. Records with missing exposure data are excluded. In settings with complete exposure and outcome data, the MCMI approach can be expected to yield an unbiased ACE estimate under an extended exchangeability assumption ($Y^x \coprod X|Z, M_Z$ for $x = 0,1$, where $M_Z$ is the vector of missingness indicators for the incomplete confounders), and the assumption that the exposure or outcome depend on the confounder only when the confounder is observed.(22, 23) This assumption is plausible in some settings, such as electronic health record data where the decision to prescribe a medication might be influenced by family history of disease only when the clinician has the relevant information.

***MI approaches***

The MI fully conditional specification (FCS) framework (24) enables simultaneous handling of missing exposure, confounder, and outcome data. Under this approach, univariate imputation models are specified for each incomplete variable conditional on other variables and imputations are drawn sequentially until convergence.(24) The process is repeated multiple times to generate multiple completed datasets. Analysis is performed within each completed dataset and the results are pooled using Rubin's rules to obtain the final estimate and its standard error (SE).(24) For valid inference with MI, each univariate imputation model should be tailored to be compatible with the analysis method. To achieve this, all analysis variables and complexities such as interaction terms in the target analysis should be included as predictors in each univariate imputation model.(24) There are various possible implementations of MI within the FCS framework:

*Parametric MI with no interaction (MI-no int):* each univariate imputation model is based on a regression model with main-effects terms only – the default in most MI software. In the example and simulations, we considered main-effects logistic regression for the binary variables and predictive mean matching (PMM) for the continuous outcome based on a main-effects linear regression. In PMM, imputed values are drawn using the nearest observed value after fitting the regression (24) which makes it robust to misspecification of the latter, for example in the presence of nonlinear associations.(25)

*Parametric MI with two-way interactions (MI-2-way int):* each univariate imputation model is based on a regression model as above, but including two-way exposure-outcome, exposure-confounder, confounder-outcome, and confounder-confounder interactions. Interaction terms are generated within each cycle of the MI algorithm from current values of relevant variables involved in the interaction term (the so-called "passive" approach in *mice* in R).(26)

*Parametric MI with two-, three-, and four-way interactions (MI-higher int):* as above, but all univariate imputation models additionally include three- and four-way confounder-confounder interactions as predictors.

*MI using classification and regression trees (MI-CART):* instead of regression, for each variable with missing data a tree is fitted using a recursive partitioning technique, with all other variables as predictors. Each record belongs to a donor leaf, from which a randomly selected value for the variable is taken as the imputed value.(27) *MI-CART* (and *MI using random forest (MI-RF)*, see below) have been proposed to enable imputation that can more flexibly allow for interactions and non-linearities.(27)

*MI using random forest (MI-RF)*: for each variable with missing data, multiple bootstrap samples are drawn and for each a separate tree is fitted. Each tree contributes a donor leaf, and a randomly selected value for the variable from all these donors is taken as the imputed value.(27)

All MI approaches can be implemented with the *mice* package in R.(26)

## Simulation study

To compare the performance of the described methods for handling missing data, we performed a simulation study based on the VAHCS example (**Figure 1**). We considered six scenarios. For each, we generated 2,000 datasets of 2,000 records.

### *Generating the complete data*

We used parametric regression models to generate the variables. The values of parameters in the models were determined by fitting similar models to the available data in VAHCS (unless stated otherwise). We considered two data generating scenarios (*simple* and *complex*), differing in the confounder-confounder interaction terms used in the data generation models. No exposure-confounder interaction terms were included (no effect modification). **Supplementary Table 1** provides the parameter values used for simulating the data and **Supplementary Table 2** the descriptive statistics of the variables in the simulated data.

For both scenarios, we generated a continuous auxiliary variable A(age at wave 2) and a set of confounders $Z = (Z_1$(parental divorce), $Z_2$(antisocial behaviour), $Z_3$(depression and anxiety), $Z_4$(alcohol use), $Z_5$(parental education)). The models for generating these variables were (all binary variables coded 0/1 and $\text{logit}^{-1}((\cdot)) = \exp(\cdot)/(1 + \exp(\cdot))$:

$$A \sim N(0,1)$$
$$Z_1 \sim \text{Binomial}(1, \text{logit}^{-1}(\alpha_0))$$
$$Z_2 \sim \text{Binomial}\left(1, \text{logit}^{-1}(\beta_0 + \beta_1 A)\right)$$
$$Z_3 \sim \text{Binomial}(1, \text{logit}^{-1}(\gamma_0 + \gamma_1 A))$$
$$Z_4 \sim \text{Binomial}\left(1, \text{logit}^{-1}(\delta_0 + \delta_1 A)\right)$$
$$Z_5 \sim \text{Binomial}(1, \text{logit}^{-1}(\zeta_0))$$

The scenarios differed in the exposure and outcome generation models.

*Simple scenario:* we used main-effects regression models to generate a binary exposure X(frequent cannabis use) and a continuous outcome Y(log-transformed, standardized CIS-R total score):

$$X_{\text{simple}} \sim \text{Binomial}(1, \text{logit}^{-1}(\eta_0 + \eta_1 Z_1 + \eta_2 Z_2 + \eta_3 Z_3 + \eta_4 Z_4 + \eta_5 Z_5 + \eta_6 A))$$
$$Y_{\text{simple}} \sim N(\theta_0 + \theta_1 X + \theta_2 Z_1 + \theta_3 Z_2 + \theta_4 Z_3 + \theta_5 Z_4 + \theta_6 Z_5, \text{sd} = 1)$$

*Complex scenario:* we used regression models that included confounder-confounder interactions (excluding interactions with $Z_2$ because of the low prevalence (15%)) to generate the exposure and outcome:

$$\begin{aligned} X_{\text{complex}} \sim \text{Binomial}(1, \text{logit}^{-1}(&\eta_0^* + \eta_1 Z_1 + \eta_2 Z_2 + \eta_3 Z_3 + \eta_4 Z_4 + \eta_5 Z_5 + \eta_6 A + \eta_7 Z_1 Z_3 + \eta_8 Z_1 Z_4 \\ &+ \eta_9 Z_1 Z_5 + \eta_{10} Z_3 Z_4 + \eta_{11} Z_3 Z_5 + \eta_{12} Z_4 Z_5)) \end{aligned}$$
$$\begin{aligned} Y_{\text{complex}} \sim N(&\theta_0^* + \theta_1 X + \theta_2 Z_1 + \theta_3 Z_2 + \theta_4 Z_3 + \theta_5 Z_4 + \theta_6 Z_5 + \theta_7 Z_1 Z_3 + \theta_8 Z_1 Z_4 + \theta_9 Z_1 Z_5 + \theta_{10} Z_3 Z_4 \\ &+ \theta_{11} Z_3 Z_5 + \theta_{12} Z_4 Z_5 + \theta_{13} Z_1 Z_3 Z_4 + \theta_{14} Z_1 Z_3 Z_5 + \theta_{15} Z_1 Z_4 Z_5 + \theta_{16} Z_3 Z_4 Z_5 \\ &+ \theta_{17} Z_1 Z_3 Z_4 Z_5, \text{sd} = 1) \end{aligned}$$

We inflated the coefficients for the interaction terms, approximately four times larger than values estimated in the VAHCS data. Under both outcome generation models, we set the coefficients for X ($\theta_1$), i.e., the true value of the ACE, to 0.2. For this effect and 2,000 records, the null hypothesis of no causal effect is formally rejected ($p < 0.05$) in approximately 80% of the simulated datasets.

### *Imposing missing data*

We considered missingness scenarios depicted by m-DAGs (missingness directed acyclic graphs) A and B (**Figure 2**).(19, 28). These were selected as they represent scenarios under which our expectations of the performance of methods were quite distinct (see Discussion).

We imposed missingness on $Z_2, Z_3, Z_4, X, Y$ through generating missingness indicators $M_{Z_2}, M_{Z_3}, M_{Z_4}, M_X, M_Y$, coded 0 if the variable was observed; 1 if missing. We considered variables $A, Z_1, Z_5$, which had <10% missing within VAHCS (**Table 1**), as fully observed in the simulation study.

For the *simple scenario*, the models used for generating the missingness indicators according to each m-DAG were:

$$M_{Z_2} \sim \text{Binomial}\left(1, \text{logit}^{-1}(\iota_0 + \iota_1 Z_1 + \iota_2 Z_5 + \iota_3 Z_2 + \iota_4 X + \iota_5 Y)\right)$$
$$M_{Z_3} \sim \text{Binomial}(1, \text{logit}^{-1}(\kappa_0 + \kappa_1 Z_1 + \kappa_2 Z_5 + \kappa_3 Z_3 + \kappa_4 X + \kappa_5 Y + \kappa_6 M_{Z_2}))$$
$$M_{Z_4} \sim \text{Binomial}(1, \text{logit}^{-1}(\lambda_0 + \lambda_1 Z_1 + \lambda_2 Z_5 + \lambda_3 Z_4 + \lambda_4 X + \lambda_5 Y + \lambda_6 M_{Z_2} + \lambda_7 M_{Z_3}))$$

$$M_X \sim \text{Binomial}(1, \text{logit}^{-1}(\nu_0 + \nu_1 Z_1 + \nu_2 Z_5 + \nu_3 Z_2 + \nu_4 Z_3 + \nu_5 Z_4 + \nu_6 X + \nu_7 Y + \nu_8 M_{Z_2} + \nu_9 M_{Z_3} + \nu_{10} M_{Z_4}))$$
$$M_Y \sim \text{Binomial}(1, \text{logit}^{-1}(\xi_0 + \xi_1 Z_1 + \xi_2 Z_5 + \xi_3 Z_2 + \xi_4 Z_3 + \xi_5 Z_4 + \xi_6 X + \xi_7 Y + \xi_8 M_{Z_2} + \xi_9 M_{Z_3} + \xi_{10} M_{Z_4} + \xi_{11} M_X))$$

For each missingness indicator, we set the coefficients for all confounders and the exposure to 0.9. We set the coefficient for the outcome to 0 for m-DAG A and to 0.1 for m-DAG B.

For the *complex scenario*, we considered two sets of models for generating missingness according to each m-DAG. The first set, hereafter *complex scenario 1,* used the same models as for the *simple scenario*, detailed above. The second set, hereafter *complex scenario 2*, used the models:

$$M_{Z_2} \sim \text{Binomial}\big(1, \text{logit}^{-1}(\iota_0 + \iota_1 Z_1 + \iota_2 Z_5 + \iota_3 Z_2 + \iota_4 X + \iota_5 Y + \iota_6 XZ_2 + \iota_7 Y^2)\big)$$
$$M_{Z_3} \sim \text{Binomial}(1, \text{logit}^{-1}(\kappa_0 + \kappa_1 Z_1 + \kappa_2 Z_5 + \kappa_3 Z_3 + \kappa_4 X + \kappa_5 Y + \kappa_6 M_{Z_2} + \kappa_7 XZ_3 + \kappa_8 Y^2))$$
$$M_{Z_4} \sim \text{Binomial}(1, \text{logit}^{-1}(\lambda_0 + \lambda_1 Z_1 + \lambda_2 Z_5 + \lambda_3 Z_4 + \lambda_4 X + \lambda_5 Y + \lambda_6 M_{Z_2} + \lambda_7 M_{Z_3} + \lambda_8 XZ_4 + \lambda_9 Y^2))$$
$$M_X \sim \text{Binomial}(1, \text{logit}^{-1}(\nu_0 + \nu_1 Z_1 + \nu_2 Z_5 + \nu_3 Z_2 + \nu_4 Z_3 + \nu_5 Z_4 + \nu_6 X + \nu_7 Y + \nu_8 M_{Z_2} + \nu_9 M_{Z_3} + \nu_{10} M_{Z_4} + \nu_{11} XZ_2 + \nu_{12} XZ_3 + \nu_{13} XZ_4 + \nu_{14} Y^2))$$
$$M_Y \sim \text{Binomial}(1, \text{logit}^{-1}(\xi_0 + \xi_1 Z_1 + \xi_2 Z_5 + \xi_3 Z_2 + \xi_4 Z_3 + \xi_5 Z_4 + \xi_6 X + \xi_7 Y + \xi_8 M_{Z_2} + \xi_9 M_{Z_3} + \xi_{10} M_{Z_4} + \xi_{11} M_X + \xi_{12} XZ_2 + \xi_{13} XZ_3 + \xi_{14} XZ_4))$$

For the *complex scenario 2*, we set the coefficients for all confounders, the exposure, and the outcome the same values as for the *simple scenario*, detailed above. We set the coefficients for exposure-confounder interactions ($XZ_2, XZ_3, XZ_4$) to 0.9, and for $Y^2$ to 0 for m-DAG A and 0.08 for m-DAG B.

This led to overall 6 scenarios (2 m-DAGs × 3 data generating scenarios). The missingness proportions were the same in all missingness scenarios and approximately the same as in the real VAHCS dataset, except for the outcome, which was increased to 20% (13% in VAHCS).

The overlap between missingness proportions were adjusted so that the proportions of records excluded were 50% for *CCA*, 40% for *Ext-TMLE+CEC*, and 30% for *Ext-TMLE+MCMI* (**Supplementary Table 2**).

***Analysis of the simulated data***

The target analysis aimed to estimate the ACE of X on Y using TMLE with data-adaptive methods adjusting for $Z_1, Z_2, Z_3, Z_4, Z_5$ as confounders. We used the *TMLE* package in R (20). We fitted the exposure and outcome models using a SuperLearner library that included a range of parametric, semiparametric, and non-parametric methods.(29, 30) The SE for TMLE was obtained using the variance of the influence function.(9) The analysis was applied to each simulated incomplete dataset alongside each of the previously described missing data methods. For *Ext-TMLE+CEC* and *Ext-TMLE+MCMI* we used the same SuperLearner library for the outcome missingness model. We used the *mice* package in R to implement MI.(26) Due to computational constraints, for each MI approach, we generated five imputed datasets (see Discussion).(24)

**Supplementary Tables 3** and **4** show the variables and interaction terms included in each imputation model for *MI-2-way int* and *MI-higher int*. We used the default settings of the *mice* package for the donor pool for PMM in parametric MI approaches and for the hyperparameters for *MI-CART* and *MI-RF*.(26)

### ***Evaluation criteria***

We compared the performance of the approaches for handling missing data by calculating the relative bias percent, the empirical standard errors (SE), and the percent error in average model-based SE relative to the empirical SE. For all, Monte-Carlo SEs (MC-SE) were obtained.(31)

Analyses were performed in R version 3.6.1.(32)

### ***Simulation study results***

#### *Relative bias*

In the *simple scenario*, under m-DAG A*, CCA* and *Ext-TMLE+CEC* yielded small biases (≤3%). *Ext-TMLE+MCMI* was more biased (10%) (**Figure 3**). These three approaches led to larger biases under m-DAG B (-13% to -16%). The three parametric MI approaches performed similarly to each other under both m-DAGs, yielding small biases (<7%). *MI-CART* performed similarly to the parametric MI approaches under m-DAG A (relative bias -8%), but it had larger bias under m-DAG B (-20%). Of all the approaches, *MI-RF* had the larger bias under both m-DAGs (-32% m-DAG A, -42% m-DAG B).

Biases in *complex scenario 1*, for both m-DAGs, were similar to those in the *simple scenario*, except that *MI-no int* had larger bias than in the *simple scenario* (33% m-DAG A, 23% m-DAG B).

In the *complex scenario 2*, for m-DAG A, biases for non-MI methods (<|15%|), *MI-CART* (-24%) and *MI-RF* (-58%) were larger than in prior scenarios, while the three parametric MI approaches performed similarly to the *complex scenario 1*. For m-DAG B, all the non-MI and parametric MI approaches and *MI-RF* had larger bias than in prior scenarios (-42% to -43% for non-MI approaches, 35% to 43% for parametric MI approaches, -64% for *MI-RF*), while *MI-CART* had similar bias as in the *simple scenario* and *complex scenario 1*.

The MC-SE for relative bias ranged from 0.81%-1.98% across scenarios and m-DAGs.

#### *Empirical standard error and relative error in model-based standard error*

Across scenarios and m-DAGs, the empirical SEs using *CCA* and *Ext-TMLE+CEC* were similar (0.13-0.16) and larger than *Ext-TMLE+MCMI* (0.12-0.14) (**Figure 4**). The SEs obtained from the parametric MI approaches were similar to each other (0.10-0.14), to *Ext-TMLE+MCMI*, and *MI-CART*, except for *complex scenario 2* under m-DAG B, where the parametric MI approaches exhibited larger SEs (0.16-0.18). The empirical SEs were similar across scenarios and m-DAGs for *MI-CART* (0.10-0.13) and *MI-RF* (0.07-0.09).

*MI-RF* had the lowest empirical SE in all scenarios and m-DAGs (0.7-0.8). The MC-SE for empirical SEs ranged from 0.001-0.003.

The model SEs were underestimated using non-MI methods and overestimated using MI methods across all scenarios and m-DAGs (**Figure 5**). The errors were smallest under the *simple scenario* and largest under the *complex scenario 2*. Within each scenario, the performance among non-MI approaches was similar. The performance of the MI approaches was similar within each scenario, except *MI-RF*, which produced model SEs with considerably larger error. The MC-SE for relative % error in model-based SEs ranged from 1.15%-3.02%.

## Illustrative example results

We analysed the VAHCS example using the *tmle* package in R,(20) and applied the eight missing data methods described. Unlike in the simulations, a small proportion of participants had missing data for parental divorce and parental education (**Table 1**), which were handled in the same way as missing data for the other confounders. Also, the auxiliary variable age had 9.3% missing data, which was multiply imputed in the MI approaches. For the MI approaches, 100 imputations were performed.

The obtained effect sizes were small, with *MI-no int* yielding somewhat larger effect size and non-MI methods and *MI-RF* smaller effect sizes (**Table 2**). The SEs for MI approaches were larger than the non-MI methods, which could be explained by the downward and upward biases in model SEs for non-MI and MI approaches, respectively, observed in our simulation study (**Figure 5**). For example, using the relative percent error in model SEs averaged over the six scenarios in the simulations, the corrected SEs in the case study would be 0.13 for *CCA*, 0.13 for *MI-no int* and 0.11 for *MI-RF*.

In the VAHCS example, the outcome (mental health in young adulthood) might well have influenced its own missingness, in which case neither of the considered m-DAGs in the simulation study are plausible for our example and we would expect all the considered missing data methods to be biased.(19)

## Discussion

We compared methods for handling missing data when estimating the ACE using TMLE with data-adaptive approaches. We considered six scenarios with different exposure and outcome generation models (presence/absence of confounder-confounder interaction terms) and missingness mechanisms (whether the outcome influenced missingness in other variables and presence/absence of interaction/non-linear terms in missingness models). *CCA* and *EXT-TMLE+CEC* had small bias under m-DAG A (where the outcome did not influence missingness in other variables), and large bias otherwise. *MI-no int* – the default in most MI software – had large bias in the *complex scenarios 1* and *2* (when the exposure/outcome generation models included interactions) and small bias otherwise (regardless of m-DAG). *MI-2-way int* and *MI-higher int* performed best in terms of bias and variance across all settings, except for m-DAG B in *complex scenario 2* (where a non-linear outcome term influenced missingness in other variables). *MI-RF* had consistently large bias across the six scenarios. *MI-CART* had small bias under m-DAG A in the *simple* and *complex scenario 1*, and large bias otherwise.

Based on previous investigations in a setting without effect modification, we determined that for m-DAG A, where the outcome did not influence missingness in any variable, the ACE was identifiable (or "recoverable").(19) Further, because auxiliary variables did not influence missingness in any variable, we expected both *CCA* and an appropriate implementation of MI to yield low bias (33). Indeed, *CCA* (and *Ext-TMLE+CEC*) produced estimates with small bias for m-DAG A across all data generation scenarios. Also, implementations of parametric MI that were approximately compatible with the analysis method (i.e., all parametric MI procedures in the *simple scenario*, and MI including interaction terms in *complex scenarios*) returned estimates with little bias, while an inappropriate MI method (e.g., *MI-no int* in our complex scenarios) was considerably more biased.

Contrary to *CCA* and *Ext-TMLE+CEC,* the *Ext-TMLE+MCMI* approach had higher bias under m-DAG A. A key assumption under which the MCMI approach has been shown to be unbiased is when the exposure or outcome only depend on the confounder when the confounder is observed.(22, 23) We did not consider missingness scenarios where this held, because this assumption is implausible in a prospective cohort study, like VAHCS, where the data are not used for medical decision-making.

For m-DAG B, the ACE was determined to be non-recoverable, but since the outcome did not influence its own missingness, based on a previous simulation study (19), we speculated that an implementation of MI that was tailored to the analysis method may offer some bias reduction compared with *CCA*. We observed that parametric MI approaches including interactions performed better than all other approaches in terms of bias for the *simple scenario* and *complex scenario 1*. For the *complex scenario 2*, where the missingness models included exposure-confounder interactions and a quadratic term for the outcome, all parametric MI approaches were highly biased. *MI-CART* outperformed parametric MI approaches in this scenario, but it was still moderately biased.

Within the FCS framework, recursive partitioning techniques, such as CART and RF, have been suggested as alternative approaches that could automatically incorporate interactions and non-linearities in the imputation process.(27) Previous simulation studies have shown that MI using CART performs better than parametric MI without interaction terms.(27, 34) However, in these studies, the target analysis was a correctly specified outcome regression model with interactions, and biases in estimates of the main effects were not that different following MI using CART or parametric MI without interaction. These studies imposed missingness in the outcome only (27) or outcome and covariates.(34) In both, missingness depended on fully observed variables. In the present study, the only setting where *MI-CART* outperformed parametric MI approaches was for m-DAG B in the *complex scenario 2*. In all scenarios, bias in the ACE estimates following *MI-RF* was larger than *MI-CART*, consistent with Dove et al.'s results.(27) We speculate that this might have been because in the implementation of *MI-RF* used a small number of randomly preselected predictor variables to split at each node, which might have negatively impacted prediction accuracy.(35) Specifically, the *ranger* package used within R's *mice* package for implementing *RF* uses, as the default number of preselected variables to split at each node, the square root of the total number of variables in the imputation model, which was 2 in our simulation.(27) *CART*, on the other hand, does not involve variable preselection.(35)

In this study, non-MI approaches underestimated the model SE, which was not surprising. If both the exposure and outcome models are consistently estimated and the Donsker class condition is satisfied, TMLE is an asymptotically linear estimator and its variance can be obtained based on the variance of the influence curve.(9, 10) It is, however, unclear if the Donsker class condition is met when data-adaptive approaches are used for the exposure and outcome models.(36) This leads to bias in variance estimation as has been observed here and in other simulation studies.(30, 36, 37) It is an ongoing area of research to develop approaches to tackle it, such as cross-validated TMLE, which allows asymptotic linearity to be established without the Donsker class condition.(10) Additionally, Rubin's MI variance estimator is expected to perform poorly in the presence of incompatibility,(38) which might explain the overestimation of model SEs for the MI approaches. Incompatibility is the key challenge for using MI with TMLE with data-adaptive approaches, in terms of bias of point estimates as discussed previously, but even more so for bias in variance estimates. A promising alternative approach for obtaining SEs for MI in the presence of incompatibility has been recently proposed using the bootstrap,(38) but we did not explore this because of computational constraints. Incompatibility between the models used in MI and TMLE may also compromise the double robustness property of TMLE.(39) Other approaches, such as *EXT-TMLE* considered in this paper to handle missing outcome data, and an alternative likelihood parameterization to construct doubly robust estimators in adjusting for confounding and missing data in the presence of missing confounder data,(39) guarantee double robustness under extended assumptions.

Our simulation study was broadly based on VAHCS. We evaluated the performance of missing data methods under various missingness mechanisms. To describe what variables influenced missingness, we used m-DAGs because the standard classification of missing completely at random (MCAR), missing at random (MAR), or missing not at random (MNAR) is difficult to comprehend and substantively assess when there is missingness in multiple variables. Also, although it is possible to estimate key parameters unbiasedly if the MAR assumption holds, MAR is not necessary for unbiased estimation.(19, 33) We did not consider missingness mechanisms where outcome influenced its own missingness, under which none of the approaches could be expected to perform well. For each MI approach, due to computational constraints we generated five completed datasets in the simulation study, which is fewer than we would do in practice.(24) We do not expect this to have affected the comparison between MI approaches, but it could have affected comparison of non-MI with MI methods. Our simulated data had a relatively simple structure across the assessed scenarios. Extensions of our study could investigate the performance of these missing data methods for datasets with high-dimensional confounders, binary outcome, and more complex m-DAGs including longitudinal auxiliary variables.

## Conclusion

We evaluated the performance of eight available approaches to handle missing data when estimating the ACE using TMLE with data-adaptive approaches under various data generation scenarios and missingness mechanisms. Our results highlight the importance of considering the missingness mechanism and compatibility with the analysis method when choosing a method to handle missing data. In many settings, a parametric MI approach that incorporates interactions and non-linearities is expected to perform well in the context of TMLE with data-adaptive approaches.

**Table 1** – Description of variables used in the case study, their distribution and the proportion with missing data among the VAHCS female participants (n=1,000)

| | Variable | Type | Grouping/unit | Notation | N (%*) coded 1 or mean (SD) | % with missing data |
|---|---|---|---|---|---|---|
| Confounder | Parental divorce | Binary | 0=Not divorced/separated by wave 6<br>1= Divorced/separated by wave 6 | Z1 | 221 (22.1) | 0.1 |
| | Antisocial behaviour | Binary | 0=No across all waves 2 to 6<br>1=Yes at any wave 2 to 6 | Z2 | 106 (14.6) | 27.4 |
| | Depression and anxiety | Binary | 0=CIS-R score <12 across all waves 2 to 6<br>1=CIS-R score ≥12 at any wave 2 to 6 | Z3 | 516 (59.9) | 13.8 |
| | Alcohol use | Binary | 0=No across all waves 2 to 6<br>1=Yes at any wave 2 to 6 | Z4 | 294 (37.2) | 21.0 |
| | Parental education | Binary | 0=Did not complete high school by wave 6<br>1=Completed high school by wave 6 | Z5 | 364 (37.7) | 3.4 |
| Exposure | Frequent cannabis use | Binary | 0=Less than weekly use across all waves 2 to 6<br>1=At least weekly use at any wave 2 to 6 | X | 86 (12.4) | 30.8 |
| Outcome | CIS-R total score | Continuous | z-score, measured at wave 7 | Y | 0 (1) | 13.4 |
| Auxiliary variable | Age | Continuous | years, measured at wave 2 | A | 15.4 (0.4) | 9.3 |
| With any missing data | | | | | | 40.3 |

Abbreviations N number; SD standard deviation; CIS-R revised clinical interview schedule *Proportions reported among those with observed data for the variable

**Table 2** – Estimated average causal effect (ACE) of frequent cannabis use during adolescence on CIS-R score (standardised z-score) using a TMLE approach under different missing data methods within the VAHCS case study

| Missing data method | ACE (Difference in means)* | Standard error | 95% confidence interval | Time to run |
|---|---|---|---|---|
| Complete-case | 0.09 | 0.12 | -0.14, 0.32 | 16.4 sec |
| Ext-TMLE | 0.12 | 0.11 | -0.09, 0.33 | 11.2 sec |
| Ext-TMLE+MCMI | 0.13 | 0.13 | -0.13, 0.39 | 21.7 sec |
| MI, no int | 0.20 | 0.16 | -0.11, 0.50 | 4.6 min |
| MI, 2-way int | 0.16 | 0.17 | -0.17, 0.49 | 5.8 min |
| MI, higher int | 0.18 | 0.16 | -0.13, 0.49 | 5.8 min |
| MI, CART | 0.15 | 0.16 | -0.16, 0.45 | 11.8 min |
| MI, RF | 0.13 | 0.18 | -0.21, 0.48 | 14.1 min |

*Average causal effect estimated as the difference in the mean potential outcome under exposure and under no exposure

Abbreviations – Ext-TML: Eextended targeted maximum likelihood estimation (TMLE) approach; Ext-TMLE+MCMI: extended TMLE plus missing covariate missing indicator (MCMI) approach; MI, no int (linear): parametric multiple imputation (MI) with no interaction – linear regression to impute missing outcome; MI no int: parametric MI with no interaction – predictive mean matching to impute missing outcome; MI, 2-way int: parametric MI with two-way interactions; MI higher int: parametric MI with two-, three-, and four-way interactions; MI, CART: MI using classification and regression trees; MI, RF: MI using random forest

## Figures

**Figure 1** – Directed acyclic graph (DAG) used in data generation for the simulation study; abbreviations: CIS-R revised clinical interview schedule

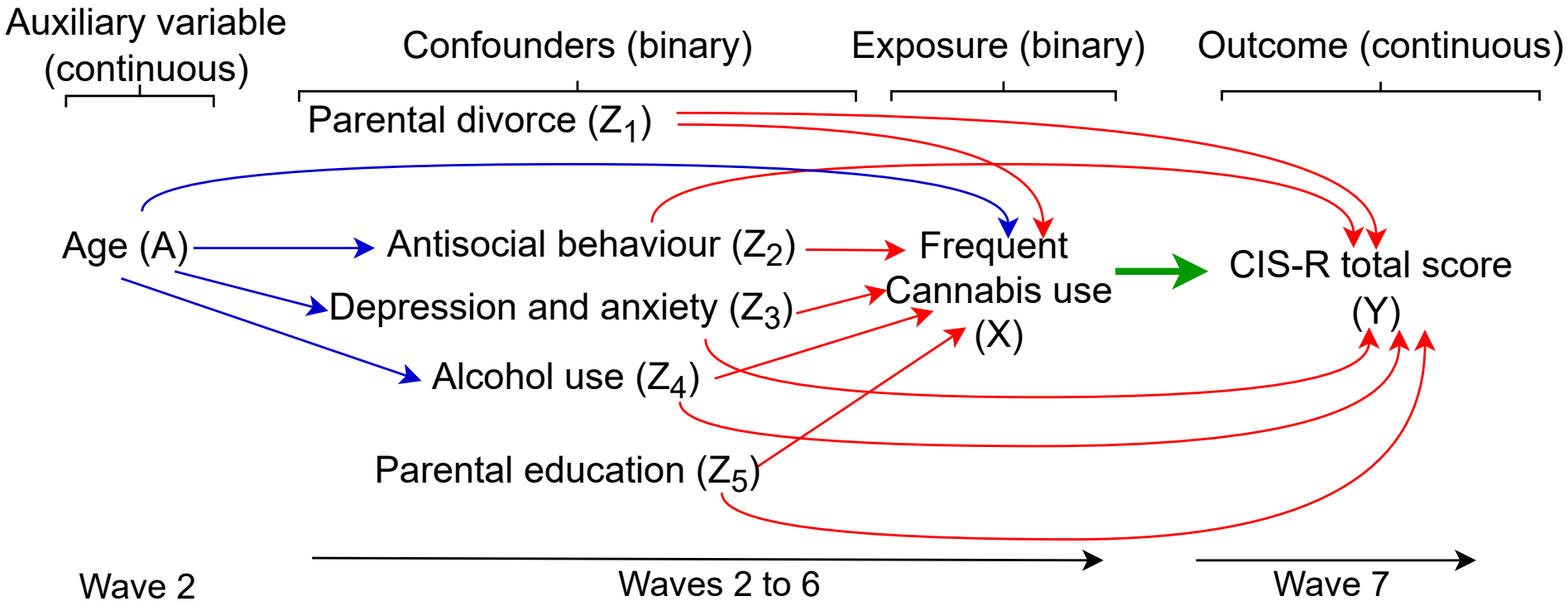

**Figure 2** – Missingness directed acyclic graphs (m-DAGs) illustrating the missingness scenarios considered in the simulation study. Figure has been adapted from Moreno-Betancur et al. [19] For simplicity of exposition, confounders without missing data ($Z_1$ and $Z_5$) are presented on a single node and confounders with missing data ($Z_2$, $Z_3$, $Z_4$) on another single node. Also, only one missingness indicator has been included for confounders with missing data ($M_Z$), coded as 1 when any of the variables $Z_2$, $Z_3$, $Z_4$ have missing data and as 0 when none has missing data.

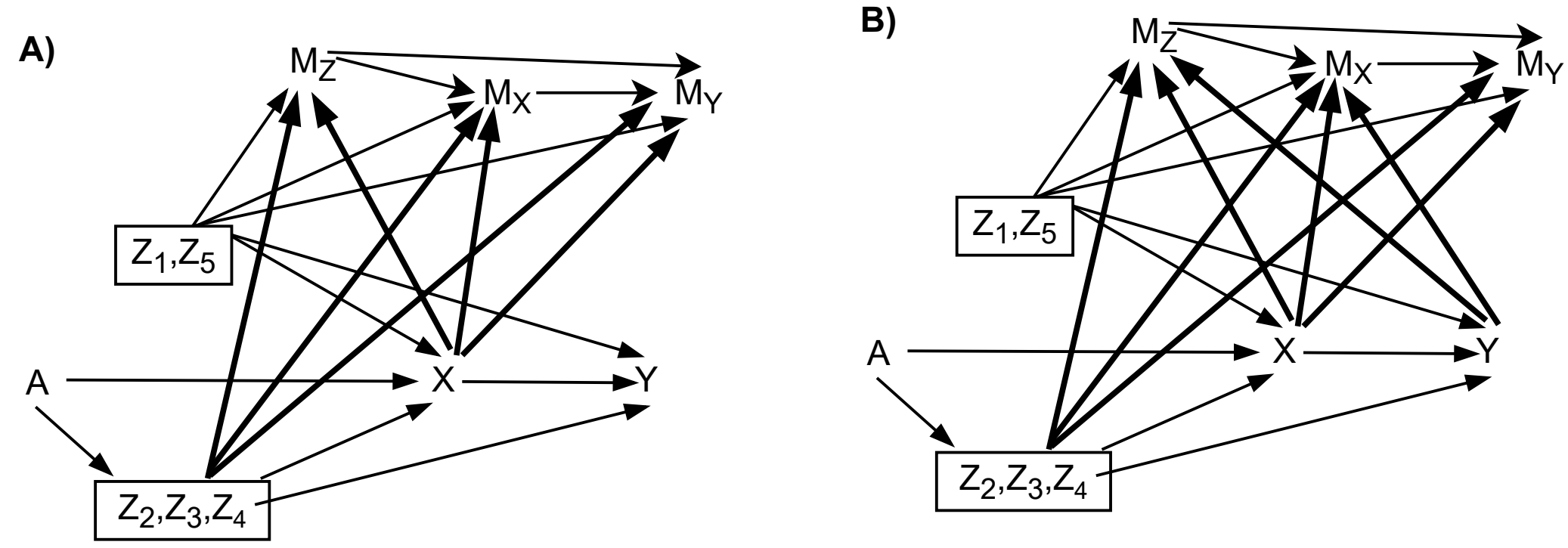

**Figure 3** – **Relative bias percent** (%) in ACE estimation (filled circles with error bars showing ±Monte Carlo standard errors) using different missing data methods for simple and complex scenarios and m-DAGs A and B. For all missing data methods, TMLE was implemented using the SuperLearner including the following methods: mean (the average), glm (generalized linear model), glm.interaction (generalized linear model with 2-way interactions between all pairs of variables), bayesglm (Bayesian generalized linear model), gam (generalized additive model), glmnet (elastic net regression), earth (multivariate adaptive regression splines), rpart (recursive partitioning and regression trees), rpartPrune (recursive partitioning with pruning) and ranger (random forest). Abbreviations – *Ext-TML*: extended targeted maximum likelihood estimation (TMLE) approach; *Ext-TMLE+MCMI*: extended TMLE plus missing covariate missing indicator (MCMI) approach; *MI-no int (linear)*: parametric multiple imputation (MI) with no interaction – linear regression to impute missing outcome; *MI-no int*: parametric MI with no interaction – predictive mean matching to impute missing outcome; *MI-2-way int*: parametric MI with two-way interactions; *MI-higher int*: parametric MI with two-, three-, and four-way interactions; *MI-CART*: MI using classification and regression trees; *MI-RF*: MI using random forest.

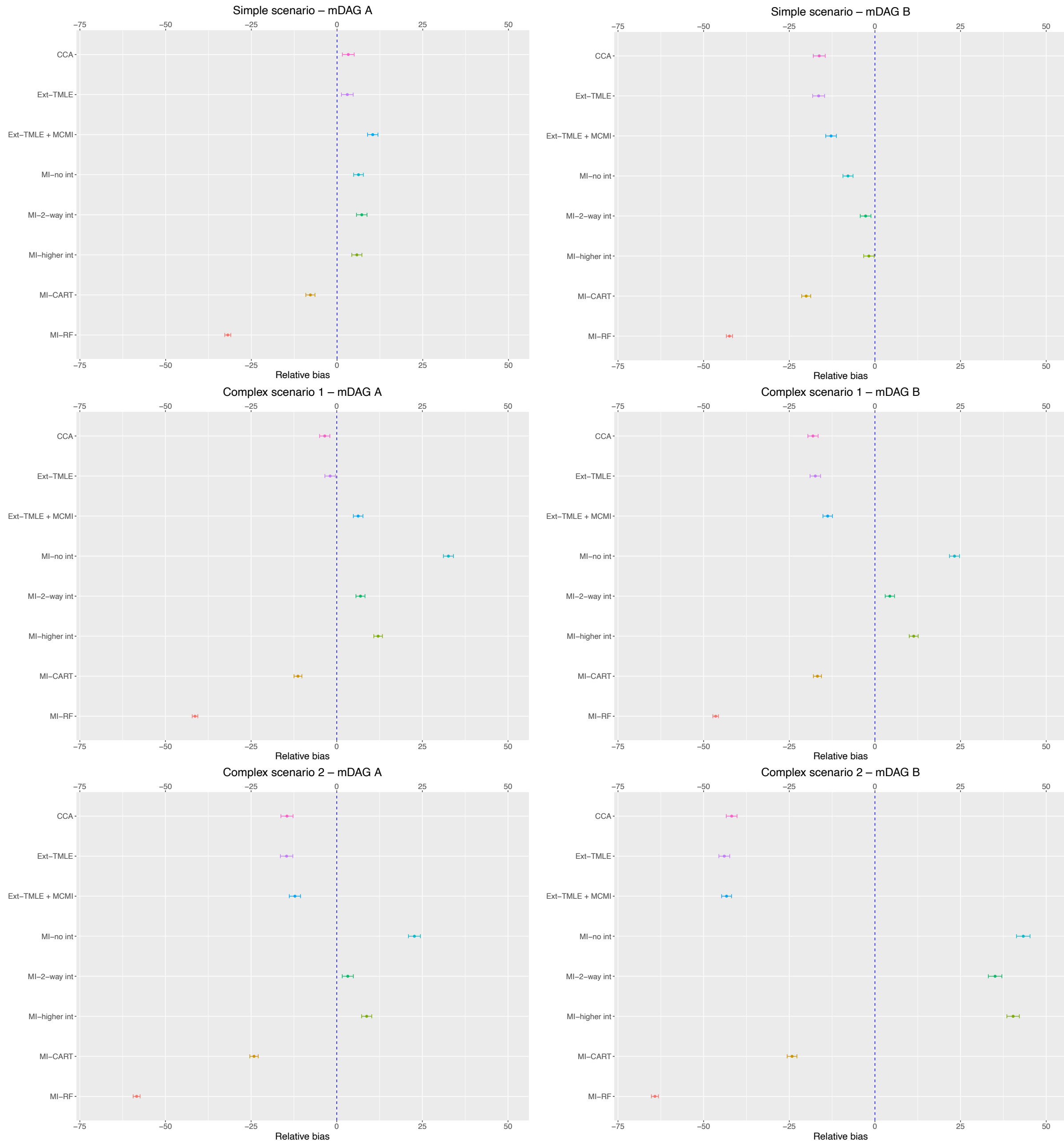

**Figure 4** – **Empirical standard error** in ACE estimation (filled circles with error bars showing ±Monte Carlo standard errors) using different missing data methods for simple and complex scenarios and m-DAGs A and B. Abbreviations – *Ext-TML*: Extended targeted maximum likelihood estimation (TMLE) approach; *Ext-TMLE+MCMI*: extended TMLE plus missing covariate missing indicator (MCMI) approach; *MI-no int (linear)*: parametric multiple imputation (MI) with no interaction – linear regression to impute missing outcome; *MI-no int*: parametric MI with no interaction – predictive mean matching to impute missing outcome; *MI-2-way int*: parametric MI with two-way interactions; *MI-higher int*: parametric MI with two-, three-, and four-way interactions; *MI-CART*: MI using classification and regression trees; *MI-RF*: MI using random forest.

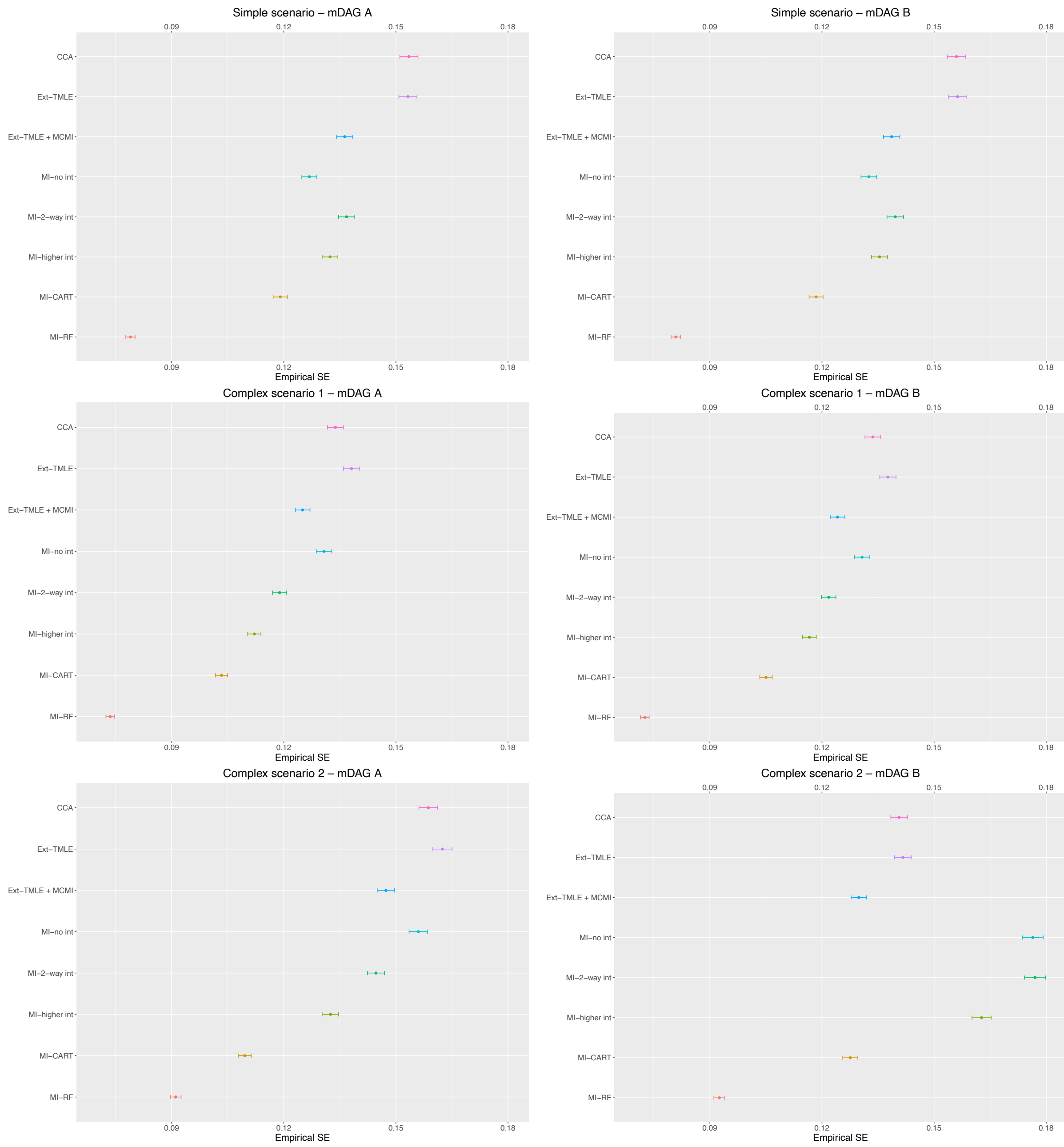

**Figure 5** – **Relative % error in model standard error** in ACE estimation (filled circles with error bars showing ±Monte Carlo standard errors) using different missing data methods for simple and complex scenarios and m-DAGs A and B. Abbreviations – *Ext-TML*: Extended targeted maximum likelihood estimation (TMLE) approach; *Ext-TMLE+MCMI*: extended TMLE plus missing covariate missing indicator (MCMI) approach; *MI-no int (linear)*: parametric multiple imputation (MI) with no interaction – linear regression to impute missing outcome; *MI-no int*: parametric MI with no interaction – predictive mean matching to impute missing outcome; *MI-2-way int*: parametric MI with two-way interactions; *MI-higher int*: parametric MI with two-, three-, and four-way interactions; *MI-CART*: MI using classification and regression trees; *MI-RF*: MI using random forest.

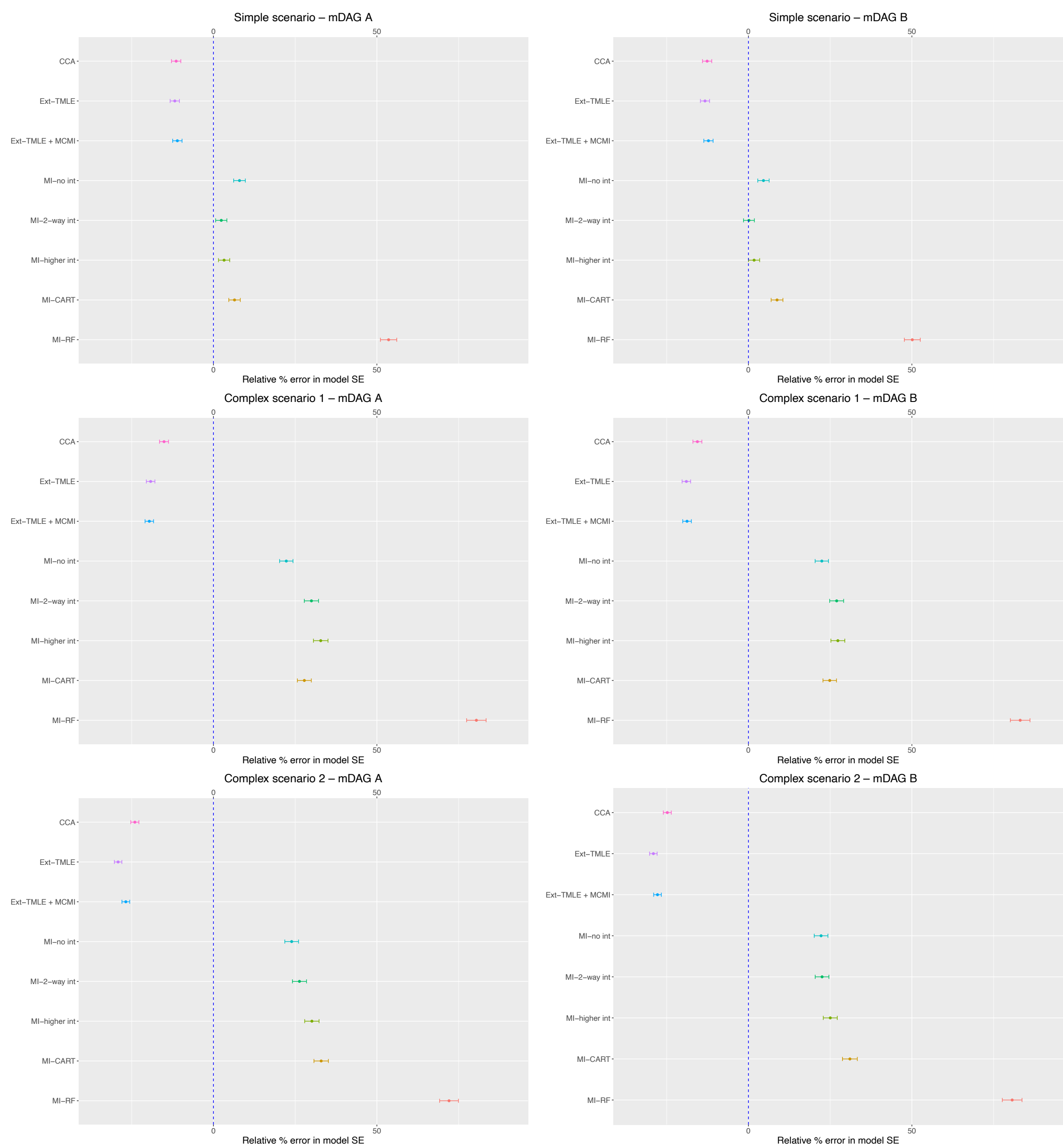

Supplementary Table 1 - Coefficent values used to simulate the variables and missingness indictors under the three assessed scenarios (simple, intermediate, and complex)

| | Model for | Regression coefficent of | | | | | | | | | | | | |
|---|---|---|---|---|---|---|---|---|---|---|---|---|---|---|
| | | Intercept | Z1 | Z2 | Z3 | Z4 | Z5 | X | Y | A | MZ2 | MZ3 | MZ4 | MX |
| | Z1 | -1.30 | | | | | | | | | | | | |
| | Z2 | -1.90 | | | | | | | | 0.40 | | | | |
| | Z3 | 0.40 | | | | | | | | 0.70 | | | | |
| | Z4 | -0.60 | | | | | | | | -0.70 | | | | |
| | Z5 | -0.50 | | | | | | | | | | | | |
| Complete data | | Simple scenario | | | | | | | | | | | | |
| | X | -2.90 | 1.30 | 1.90 | 0.40 | 0.20 | -0.30 | | | 0.70 | | | | |
| | Y | -0.70 | 0.10 | 0.40 | 0.70 | 0.20 | 0.30 | 0.20 | | | | | | |
| | | Complex scenarios | | | | | | | | | | | | |
| | X | -2.40 | 1.30 | 1.90 | 0.40 | 0.20 | -0.30 | | | | | | | |
| | Y | -0.70 | 0.10 | 0.40 | 0.70 | 0.20 | 0.30 | 0.20 | | | | | | |
| | MZ2 | -0.85 | | | | | | | | | | | | |
| | MZ3 | -4.40 | | | | | | | | | 4.30 | | | |
| DAG T | MZ4 | -4.00 | | | | | | | | | 3.90 | 1.40 | | |
| | MX | -2.00 | | | | | | | | | 1.50 | 1.50 | 1.50 | |
| | MY | -1.60 | | | | | | | | | -0.50 | 0.50 | 0.50 | 0.50 |
| | MZ2 | -1.45 | 0.90 | | | | 0.90 | | | | | | | |
| | MZ3 | -5.60 | 0.90 | | | | 0.90 | | | | 4.80 | | | |
| DAG A | MZ4 | -4.70 | 0.90 | | | | 0.90 | | | | 3.90 | 1.50 | | |
| | MX | -2.50 | 0.90 | | | | 0.90 | | | | 1.30 | 1.50 | 1.50 | |
| | MY | -2.10 | 0.90 | | | | 0.90 | | | | 0.10 | 0.50 | 0.50 | 0.70 |
| | MZ2 | -1.60 | 0.90 | | | | 0.90 | 0.90 | | | | | | |
| | MZ3 | -5.20 | 0.90 | | | | 0.90 | 0.90 | | | 4.10 | | | |
| DAG B | MZ4 | -4.40 | 0.90 | | | | 0.90 | 0.90 | | | 3.20 | 2.00 | | |
| | MX | -3.70 | 0.90 | 0.90 | 0.90 | 0.90 | 0.90 | | | | 1.80 | 1.50 | 1.50 | |
| | MY | -3.25 | 0.90 | 0.90 | 0.90 | 0.90 | 0.90 | 0.90 | | | -0.30 | 0.10 | 0.10 | 0.10 |
| | MZ2 | -1.60 | 0.90 | | | | 0.90 | 0.90 | 0.10 | | | | | |
| | MZ3 | -5.40 | 0.90 | | | | 0.90 | 0.90 | 0.10 | | 4.30 | | | |
| DAG C | MZ4 | -4.50 | 0.90 | | | | 0.90 | 0.90 | 0.10 | | 3.50 | 1.30 | | |
| | MX | -3.70 | 0.90 | 0.90 | 0.90 | 0.90 | 0.90 | | 0.10 | | 1.70 | 1.50 | 1.50 | |
| | MY | -3.25 | 0.90 | 0.90 | 0.90 | 0.90 | 0.90 | 0.90 | | | -0.40 | 0.10 | 0.10 | 0.10 |
| | MZ2 | -1.60 | 0.90 | 0.90 | | | 0.90 | | | | | | | |
| | MZ3 | -6.20 | 0.90 | | 0.90 | | 0.90 | | | | 4.80 | | | |
| DAG D | MZ4 | -5.00 | 0.90 | | | 0.90 | 0.90 | 0.90 | | | 3.90 | 1.50 | | |
| | MX | -2.65 | 0.90 | | | | 0.90 | | | | 1.30 | 1.50 | 1.50 | |
| | MY | -2.10 | 0.90 | | | | 0.90 | | | | 0.10 | 0.10 | 0.10 | 0.10 |
| | MZ2 | -1.75 | 0.90 | 0.90 | | | 0.90 | 0.90 | | | | | | |
| | MZ3 | -5.70 | 0.90 | | 0.90 | | 0.90 | 0.90 | | | 4.10 | | | |
| DAG E | MZ4 | -4.80 | 0.90 | | | 0.90 | 0.90 | 0.90 | | | 3.20 | 2.00 | | |
| | MX | -3.80 | 0.90 | 0.90 | 0.90 | 0.90 | 0.90 | 0.90 | | | 1.50 | 1.50 | 1.50 | |
| | MY | -3.20 | 0.90 | 0.90 | 0.90 | 0.90 | 0.90 | 0.90 | | | -0.60 | 0.10 | 0.10 | 0.10 |
| | MZ2 | -1.60 | 0.90 | 0.90 | | | 0.90 | | 0.10 | | | | | |
| | MZ3 | -6.60 | 0.90 | | 0.90 | | 0.90 | | 0.10 | | 5.20 | | | |
| DAG F | MZ4 | -5.40 | 0.90 | | | 0.90 | 0.90 | | 0.10 | | 4.20 | 1.70 | | |
| | MX | -2.55 | 0.90 | | | | 0.90 | 0.90 | 0.10 | | 1.20 | 1.30 | 1.30 | |
| | MY | -2.10 | 0.90 | | | | 0.90 | | | | -0.30 | 0.10 | 0.10 | 0.40 |
| | MZ2 | -1.60 | 0.90 | | | | 0.90 | 0.90 | | | | | | |
| | MZ3 | -5.20 | 0.90 | | | | 0.90 | 0.90 | | | 4.10 | | | |
| DAG G | MZ4 | -4.45 | 0.90 | | | | 0.90 | 0.90 | | | 3.20 | 2.00 | | |
| | MX | -3.70 | 0.90 | 0.90 | 0.90 | 0.90 | 0.90 | | | | 1.70 | 1.50 | 1.50 | |
| | MY | -3.30 | 0.90 | 0.90 | 0.90 | 0.90 | 0.90 | 0.90 | 0.10 | | -0.30 | 0.10 | 0.10 | 0.10 |
| | MZ2 | -1.60 | 0.90 | | | | 0.90 | 0.90 | 0.10 | | | | | |
| | MZ3 | -5.40 | 0.90 | | | | 0.90 | 0.90 | 0.10 | | 4.30 | | | |
| DAG H | MZ4 | -4.50 | 0.90 | | | | 0.90 | 0.90 | 0.10 | | 3.50 | 1.30 | | |
| | MX | -3.65 | 0.90 | 0.90 | 0.90 | 0.90 | 0.90 | | 0.10 | | 1.50 | 1.50 | 1.50 | |
| | MY | -3.30 | 0.90 | 0.90 | 0.90 | 0.90 | 0.90 | 0.90 | 0.10 | | -0.50 | 0.30 | 3.00 | 0.10 |
| | MZ2 | -1.75 | 0.90 | 0.90 | | | 0.90 | 0.90 | 0.10 | | | | | |
| | MZ3 | -5.95 | 0.90 | | 0.90 | | 0.90 | 0.90 | 0.10 | | 4.30 | | | |
| DAG I | MZ4 | -4.80 | 0.90 | | | 0.90 | 0.90 | 0.90 | 0.10 | | 3.50 | 1.30 | | |
| | MX | -3.80 | 0.90 | 0.90 | 0.90 | 0.90 | 0.90 | 0.90 | 0.10 | | 1.50 | 1.50 | 1.50 | |
| | MY | -3.20 | 0.90 | 0.90 | 0.90 | 0.90 | 0.90 | 0.90 | | | -0.60 | 0.10 | 0.10 | 0.20 |
| | MZ2 | -1.75 | 0.90 | 0.90 | | | 0.90 | 0.90 | 0.10 | | | | | |
| | MZ3 | -5.95 | 0.90 | | 0.90 | | 0.90 | 0.90 | 0.10 | | 4.30 | | | |
| DAG J | MZ4 | -4.85 | 0.90 | | | 0.90 | 0.90 | 0.90 | 0.10 | | 3.50 | 1.30 | | |
| | MX | -3.80 | 0.90 | 0.90 | 0.90 | 0.90 | 0.90 | 0.90 | 0.10 | | 1.50 | 1.50 | 1.50 | |
| | MY | -3.20 | 0.90 | 0.90 | 0.90 | 0.90 | 0.90 | 0.90 | 0.10 | | 0.60 | 0.10 | 0.10 | 0.10 |

* For the complex scenarios models for X and Y also included interactions as follows:

| | z1z3 | z1z4 | z1z5 | z3z4 | z3z5 | z4z5 | z1z3z4 | z1z3z5 | z1z4z5 | z3z4z5 | z1z3z4z5 |
|---|---|---|---|---|---|---|---|---|---|---|---|
| | Complex scenario 1 | | | | | | | | | | |
| X | -1.60 | -1.20 | -0.50 | -0.60 | 0.30 | -1.50 | | | | | |
| Y | -0.50 | 1.00 | 0.10 | 0.10 | 0.40 | -0.10 | -1.20 | -1.00 | -0.10 | -0.40 | 1.70 |
| | Complex scenario 2 | | | | | | | | | | |
| X | -3.20 | -2.30 | -1.00 | -1.20 | 0.50 | -2.90 | | | | | |
| Y | -0.90 | 2.00 | 0.10 | 0.20 | 0.70 | -0.20 | -2.40 | -2.00 | -0.30 | -0.80 | 3.40 |

Supplementaty Table 2 - Distribution of variables in the simulated complete data and prportion with missingness observed in the real VAHCS data and across the 11 simulated missingness mechanisms and the three scenarios considered

| | | **% in simulated complete datasets** | | | | | | |
|---|---|---|---|---|---|---|---|---|
| | **Z1** | **Z2** | **Z3** | **Z4** | **Z5** | **X** | **Y** | |
| | 21 | 14 | 59 | 37 | 38 | 15 | 15 | 0 (1) |
| | | **% with missing value** | | | | | | |
| | | **Z2** | **Z3** | **Z4** | **X** | **Y** | **Z/X** | **Any** |
| Observed | | 27 | 14 | 21 | 31 | 13 | 34 | 40 |
| | **DAG T** | 30 | 15 | 20 | 30 | 20 | 40 | 50 |
| | **DAG A** | 30 | 15 | 20 | 30 | 20 | 40 | 50 |
| | **DAG B** | 30 | 15 | 20 | 30 | 20 | 40 | 50 |
| | **DAG C** | 30 | 15 | 20 | 30 | 20 | 40 | 50 |
| | **DAG D** | 30 | 15 | 20 | 30 | 20 | 40 | 50 |
| Simple scenario | **DAG E** | 30 | 15 | 20 | 30 | 20 | 40 | 50 |
| | **DAG F** | 30 | 15 | 20 | 30 | 20 | 40 | 50 |
| | **DAG G** | 30 | 15 | 20 | 30 | 20 | 40 | 50 |
| | **DAG H** | 30 | 15 | 20 | 30 | 20 | 40 | 50 |
| | **DAG I** | 30 | 15 | 20 | 30 | 20 | 40 | 50 |
| | **DAG J** | 30 | 15 | 20 | 30 | 20 | 40 | 50 |
| | **DAG T** | 30 | 15 | 20 | 30 | 20 | 40 | 50 |
| | **DAG A** | 30 | 15 | 20 | 30 | 20 | 40 | 50 |
| | **DAG B** | 30 | 15 | 20 | 30 | 20 | 40 | 50 |
| | **DAG C** | 30 | 15 | 20 | 30 | 20 | 40 | 50 |
| | **DAG D** | 30 | 15 | 20 | 30 | 20 | 40 | 50 |
| Complex scenario 1 | **DAG E** | 30 | 15 | 20 | 30 | 20 | 40 | 50 |
| | **DAG F** | 30 | 15 | 20 | 30 | 20 | 40 | 50 |
| | **DAG G** | 30 | 15 | 20 | 30 | 20 | 40 | 50 |
| | **DAG H** | 30 | 15 | 20 | 30 | 20 | 40 | 50 |
| | **DAG I** | 30 | 15 | 20 | 30 | 20 | 40 | 50 |
| | **DAG J** | 30 | 15 | 20 | 30 | 20 | 40 | 50 |
| | **DAG T** | 30 | 15 | 20 | 30 | 20 | 40 | 50 |
| | **DAG A** | 30 | 15 | 20 | 30 | 20 | 40 | 50 |
| | **DAG B** | 30 | 15 | 20 | 30 | 20 | 40 | 50 |
| | **DAG C** | 30 | 15 | 20 | 30 | 20 | 40 | 50 |
| | **DAG D** | 30 | 15 | 20 | 30 | 20 | 40 | 50 |
| Complex scenario 2 | **DAG E** | 30 | 15 | 20 | 30 | 20 | 40 | 50 |
| | **DAG F** | 30 | 15 | 20 | 30 | 20 | 40 | 50 |
| | **DAG G** | 30 | 15 | 20 | 30 | 20 | 40 | 50 |
| | **DAG H** | 30 | 15 | 20 | 30 | 20 | 40 | 50 |
| | **DAG I** | 30 | 15 | 20 | 30 | 20 | 40 | 50 |
| | **DAG J** | 30 | 15 | 20 | 30 | 20 | 40 | 50 |

Supplementary Table 3 - Simulation study results for complete data for simple, intermediate, and complex scenario, using outcome regression, g-computation including all two-way confounder-confounder interactions, and TMLE

| | Estimated mean difference | Absolute bias | | Relative bias (%) | Coverage | | Bias eliminated coverage | | Empirical SE | | Model SE | | Relative % error in model SE | | Time to run |
|---|---|---|---|---|---|---|---|---|---|---|---|---|---|---|---|
| | | Estimate | MC SE | | Estimate | MC SE | Estimate | MC SE | Estimate | MC SE | Estimate | MC SE | Estimate | MC SE | |
| **Simple scenario** | | | | | | | | | | | | | | | |
| 1 | 0.20 | 0.00 | 0.00 | 0.25 | 0.93 | 0.01 | 0.93 | 0.01 | 0.08 | 0.00 | 0.08 | 0.00 | -5.39 | 1.50 | 8.91 |
| **2** | **0.20** | **0.00** | **0.00** | **0.29** | **0.94** | **0.01** | **0.93** | **0.01** | **0.08** | **0.00** | **0.08** | **0.00** | **-5.76** | **1.49** | **8.71** |
| 3 | 0.21 | 0.01 | 0.00 | 3.63 | 0.91 | 0.01 | 0.91 | 0.01 | 0.10 | 0.00 | 0.08 | 0.00 | -17.93 | 1.30 | 10.55 |
| 4 | 0.20 | 0.00 | 0.00 | 0.27 | 0.93 | 0.01 | 0.93 | 0.01 | 0.08 | 0.00 | 0.08 | 0.00 | -5.61 | 1.49 | 14.28 |
| 5 | 0.20 | 0.00 | 0.00 | 0.27 | 0.94 | 0.01 | 0.94 | 0.01 | 0.08 | 0.00 | 0.08 | 0.00 | -4.33 | 1.51 | 29.94 |
| 6 | 0.20 | 0.00 | 0.00 | 0.25 | 0.94 | 0.01 | 0.94 | 0.01 | 0.08 | 0.00 | 0.08 | 0.00 | -4.19 | 1.52 | 3.88 |
| **7** | **0.20** | **0.00** | **0.00** | **0.30** | **0.93** | **0.01** | **0.94** | **0.01** | **0.08** | **0.00** | **0.08** | **0.00** | **-5.02** | **1.50** | **4.58** |
| 8 | 0.20 | 0.00 | 0.00 | 0.30 | 0.94 | 0.01 | 0.94 | 0.01 | 0.08 | 0.00 | 0.08 | 0.00 | -3.94 | 1.52 | 7.67 |
| 9 | 0.20 | 0.00 | 0.00 | 0.24 | 0.94 | 0.01 | 0.94 | 0.01 | 0.08 | 0.00 | 0.08 | 0.00 | -4.10 | 1.52 | 1.74 |
| **Complex scenario 1** | | | | | | | | | | | | | | | |
| 1 | 0.21 | 0.01 | 0.00 | 3.11 | 0.92 | 0.01 | 0.92 | 0.01 | 0.08 | 0.00 | 0.08 | 0.00 | -6.50 | 1.48 | 16.00 |
| **2** | **0.20** | **0.00** | **0.00** | **1.98** | **0.93** | **0.01** | **0.93** | **0.01** | **0.08** | **0.00** | **0.08** | **0.00** | **-5.71** | **1.49** | **11.05** |
| 3 | 0.22 | 0.02 | 0.00 | 7.99 | 0.91 | 0.01 | 0.91 | 0.01 | 0.09 | 0.00 | 0.08 | 0.00 | -11.40 | 1.40 | 13.62 |
| 4 | 0.20 | 0.00 | 0.00 | 2.04 | 0.93 | 0.01 | 0.93 | 0.01 | 0.08 | 0.00 | 0.08 | 0.00 | -5.50 | 1.49 | 27.69 |
| 5 | 0.20 | 0.00 | 0.00 | 2.16 | 0.93 | 0.01 | 0.93 | 0.01 | 0.08 | 0.00 | 0.08 | 0.00 | -5.09 | 1.50 | 26.22 |
| 6 | 0.21 | 0.01 | 0.00 | 4.19 | 0.92 | 0.01 | 0.93 | 0.01 | 0.08 | 0.00 | 0.08 | 0.00 | -5.81 | 1.49 | 4.91 |
| **7** | **0.21** | **0.01** | **0.00** | **2.59** | **0.93** | **0.01** | **0.93** | **0.01** | **0.08** | **0.00** | **0.08** | **0.00** | **-5.20** | **1.50** | **5.86** |
| 8 | 0.21 | 0.01 | 0.00 | 4.26 | 0.93 | 0.01 | 0.93 | 0.01 | 0.08 | 0.00 | 0.08 | 0.00 | -4.69 | 1.51 | 25.17 |
| 9 | 0.21 | 0.01 | 0.00 | 4.00 | 0.92 | 0.01 | 0.93 | 0.01 | 0.08 | 0.00 | 0.08 | 0.00 | -5.63 | 1.49 | 2.49 |
| **Complex scenario 2** | | | | | | | | | | | | | | | |
| 1 | 0.22 | 0.02 | 0.00 | 8.71 | 0.85 | 0.01 | 0.86 | 0.01 | 0.10 | 0.00 | 0.07 | 0.00 | -23.45 | 1.21 | 12.98 |
| **2** | **0.20** | **0.00** | **0.00** | **0.52** | **0.87** | **0.01** | **0.87** | **0.01** | **0.09** | **0.00** | **0.07** | **0.00** | **-21.07** | **1.25** | **12.45** |
| 3 | 0.22 | 0.02 | 0.00 | 8.03 | 0.84 | 0.01 | 0.85 | 0.01 | 0.10 | 0.00 | 0.08 | 0.00 | -25.90 | 1.17 | 15.77 |
| 4 | 0.20 | 0.00 | 0.00 | 0.13 | 0.87 | 0.01 | 0.87 | 0.01 | 0.10 | 0.00 | 0.08 | 0.00 | -21.10 | 1.25 | 32.21 |
| 5 | 0.20 | 0.00 | 0.00 | 0.64 | 0.87 | 0.01 | 0.87 | 0.01 | 0.09 | 0.00 | 0.07 | 0.00 | -21.30 | 1.24 | 23.52 |
| 6 | 0.24 | 0.04 | 0.00 | 17.66 | 0.82 | 0.01 | 0.85 | 0.01 | 0.11 | 0.00 | 0.08 | 0.00 | -24.32 | 1.20 | 6.48 |
| **7** | **0.20** | **0.00** | **0.00** | **0.67** | **0.87** | **0.01** | **0.87** | **0.01** | **0.10** | **0.00** | **0.08** | **0.00** | **-20.71** | **1.25** | **7.04** |
| 8 | 0.23 | 0.03 | 0.00 | 14.42 | 0.84 | 0.01 | 0.85 | 0.01 | 0.10 | 0.00 | 0.08 | 0.00 | -23.52 | 1.21 | 35.28 |
| 9 | 0.24 | 0.04 | 0.00 | 17.77 | 0.82 | 0.01 | 0.85 | 0.01 | 0.11 | 0.00 | 0.08 | 0.00 | -24.45 | 1.19 | 3.50 |

Abbreviations MC SE Monte Carlo standard error

Libraries were:

lib1 <-c("SL.mean","SL.glm","SL.glm.interaction", "SL.bayesglm", "SL.gam", "SL.glmnet", "SL.earth","SL.rpart", "SL.ranger")

**lib2 <-c("SL.mean","SL.glm","SL.glm.interaction", "SL.bayesglm", "SL.gam", "SL.glmnet", "SL.earth","SL.rpart", "SL.rpartPrune", "SL.ranger")**

lib3 <-c("SL.mean","SL.glm","SL.glm.interaction", "SL.bayesglm", "SL.gam", "SL.glmnet", "SL.earth","SL.rpart", "SL.rpartPrune", "SL.ranger","SL.nnet")

lib4 <-c("SL.mean","SL.glm","SL.glm.interaction", "SL.bayesglm", "SL.gam", "SL.glmnet", "SL.earth","SL.rpart", "SL.rpartPrune", "SL.ranger","SL.step", "SL.step.interaction")

lib5 <-c("SL.mean","SL.glm","SL.glm.interaction", "SL.bayesglm", "SL.gam", "SL.glmnet", "SL.earth","SL.rpart", "SL.rpartPrune", "SL.ranger","SL.randomForest")

lib6 <-c("SL.mean","SL.glm","SL.glm.interaction", "SL.bayesglm", "SL.gam", "SL.glmnet", "SL.earth")

**lib7 <-c("SL.mean","SL.glm","SL.glm.interaction", "SL.bayesglm", "SL.gam", "SL.glmnet", "SL.earth", "SL.rpartPrune")**

lib8<-c('SL.glm', 'SL.step.interaction', 'SL.earth', 'SL.mean')

lib9<-c('SL.glm', 'SL.glm.interaction', 'SL.earth', 'SL.mean')

Supplementary Table 4 - The variables and interaction terms included in each imputation model for a multiple imputation approach that included all two-way interactions

| Variable imputed | Variables included in imputation model | | | | | | | | | | | | | | | | | | | | | | |
|---|---|---|---|---|---|---|---|---|---|---|---|---|---|---|---|---|---|---|---|---|---|---|---|
| | A | Z1 | Z2 | Z3 | Z4 | Z5 | X | Y | XY | XZ1 | YZ1 | XZ3 | YZ3 | XZ4 | YZ4 | XZ5 | YZ5 | Z1Z3 | Z1Z4 | Z1Z5 | Z3Z4 | Z3Z5 | Z4Z5 |
| Z2 | 1 | 1 | 0 | 1 | 1 | 1 | 1 | 1 | 1 | 1 | 1 | 1 | 1 | 1 | 1 | 1 | 1 | 1 | 1 | 1 | 1 | 1 | 1 |
| Z3 | 1 | 1 | 1 | 0 | 1 | 1 | 1 | 1 | 1 | 1 | 1 | 0 | 0 | 1 | 1 | 1 | 1 | 0 | 1 | 1 | 0 | 0 | 1 |
| Z4 | 1 | 1 | 1 | 1 | 0 | 1 | 1 | 1 | 1 | 1 | 1 | 1 | 1 | 0 | 0 | 1 | 1 | 1 | 0 | 1 | 0 | 1 | 0 |
| X | 1 | 1 | 1 | 1 | 1 | 1 | 0 | 1 | 0 | 0 | 1 | 0 | 1 | 0 | 1 | 0 | 1 | 1 | 1 | 1 | 1 | 1 | 1 |
| Y | 1 | 1 | 1 | 1 | 1 | 1 | 1 | 0 | 0 | 1 | 0 | 1 | 0 | 1 | 0 | 1 | 0 | 1 | 1 | 1 | 1 | 1 | 1 |

Supplementary Table 5 - The variables and interaction terms included in each imputation model for a multiple imputation approach that included all two-, three-, and four-way interactions

| Variable imputed | Variables included in imputation model | | | | | | | | | | | | | | | | | | | | | | | | | | | |
|---|---|---|---|---|---|---|---|---|---|---|---|---|---|---|---|---|---|---|---|---|---|---|---|---|---|---|---|---|
| | A | Z1 | Z2 | Z3 | Z4 | Z5 | X | Y | XY | XZ1 | YZ1 | XZ3 | YZ3 | XZ4 | YZ4 | XZ5 | YZ5 | Z1Z3 | Z1Z4 | Z1Z5 | Z3Z4 | Z3Z5 | Z4Z5 | Z1Z3Z4 | Z1Z3Z5 | Z1Z4Z5 | Z3Z4Z5 | Z1Z3Z4Z5 |
| Z2 | 1 | 1 | 0 | 1 | 1 | 1 | 1 | 1 | 1 | 1 | 1 | 1 | 1 | 1 | 1 | 1 | 1 | 1 | 1 | 1 | 1 | 1 | 1 | 1 | 1 | 1 | 1 | 1 |
| Z3 | 1 | 1 | 1 | 0 | 1 | 1 | 1 | 1 | 1 | 1 | 1 | 0 | 0 | 1 | 1 | 1 | 1 | 0 | 1 | 1 | 0 | 0 | 1 | 0 | 0 | 1 | 0 | 0 |
| Z4 | 1 | 1 | 1 | 1 | 0 | 1 | 1 | 1 | 1 | 1 | 1 | 1 | 1 | 0 | 0 | 1 | 1 | 1 | 0 | 1 | 0 | 1 | 0 | 0 | 1 | 0 | 0 | 0 |
| X | 1 | 1 | 1 | 1 | 1 | 1 | 0 | 1 | 0 | 0 | 1 | 0 | 1 | 0 | 1 | 0 | 1 | 1 | 1 | 1 | 1 | 1 | 1 | 1 | 1 | 1 | 1 | 1 |
| Y | 1 | 1 | 1 | 1 | 1 | 1 | 1 | 0 | 0 | 1 | 0 | 1 | 0 | 1 | 0 | 1 | 0 | 1 | 1 | 1 | 1 | 1 | 1 | 1 | 1 | 1 | 1 | 1 |

Supplementary Table 6 - Simulation study results for complete data for simple, intermediate, and complex scenario, using outcome regression, g-computation including all two-way confounder-confounder interactions, and TMLE

| | Estimated mean difference | Absolute bias | | Relative bias (%) | Coverage | | Bias eliminated coverage | | Empirical SE | | Model SE | | Relative % error in model SE | |
|---|---|---|---|---|---|---|---|---|---|---|---|---|---|---|
| | | Estimate | MC SE | | Estimate | MC SE | Estimate | MC SE | Estimate | MC SE | Estimate | MC SE | Estimate | MC SE |
| **Simple scenario** | | | | | | | | | | | | | | |
| Outcome regression | 0.20 | 0.00 | 0.00 | 0.03 | 0.95 | 0.00 | 0.95 | 0.00 | 0.07 | 0.00 | 0.07 | 0.00 | 0.06 | 1.58 |
| g-computation | 0.20 | 0.00 | 0.00 | 0.02 | 0.95 | 0.00 | 0.95 | 0.00 | 0.07 | 0.00 | 0.07 | 0.00 | 0.41 | 1.59 |
| TMLE | 0.20 | 0.00 | 0.00 | 0.21 | 0.93 | 0.01 | 0.93 | 0.01 | 0.08 | 0.00 | 0.08 | 0.00 | -5.70 | 1.49 |
| **Complex scenario 1** | | | | | | | | | | | | | | |
| Outcome regression | 0.26 | 0.06 | 0.00 | 31.63 | 0.87 | 0.01 | 0.96 | 0.00 | 0.07 | 0.00 | 0.07 | 0.00 | 4.30 | 1.65 |
| g-computation | 0.20 | 0.00 | 0.00 | 0.46 | 0.96 | 0.00 | 0.96 | 0.00 | 0.07 | 0.00 | 0.07 | 0.00 | 4.01 | 1.65 |
| TMLE | 0.20 | 0.00 | 0.00 | 1.99 | 0.93 | 0.01 | 0.93 | 0.01 | 0.08 | 0.00 | 0.08 | 0.00 | -5.68 | 1.49 |
| **Complex scenario 2** | | | | | | | | | | | | | | |
| Outcome regression | 0.41 | 0.21 | 0.00 | 106.63 | 0.22 | 0.01 | 0.96 | 0.00 | 0.08 | 0.00 | 0.08 | 0.00 | 3.38 | 1.64 |
| g-computation | 0.18 | -0.02 | 0.00 | -8.34 | 0.93 | 0.01 | 0.94 | 0.01 | 0.07 | 0.00 | 0.07 | 0.00 | -1.66 | 1.56 |
| TMLE | 0.20 | 0.00 | 0.00 | 0.51 | 0.87 | 0.01 | 0.87 | 0.01 | 0.09 | 0.00 | 0.07 | 0.00 | -21.06 | 1.25 |

Abbreviations MC SE Monte Carlo standard error

Supplementary Table 7- Performance of missing data methods for the 11 assessed missingness mechanisms under the simple scenario

| Causal Diagram | Missing data method | Estimated mean difference | Absolute bias Estimate | Absolute bias MC SE | Relative bias (%) | Coverage (%) Estimate | Coverage (%) MC SE | Bias eliminated coverage (%) Estimate | Bias eliminated coverage (%) MC SE | Empirical SE Estimate | Empirical SE MC SE | Model SE Estimate | Model SE MC SE | Relative % error in model SE Estimate | Relative % error in model SE MC SE |
|---|---|---|---|---|---|---|---|---|---|---|---|---|---|---|---|
| | Complete data | 0.20 | 0.00 | 0.00 | 0.27 | 93.20 | 0.56 | 93.20 | 0.56 | 0.08 | 0.00 | 0.08 | 0.00 | -5.61 | 1.51 |
| Causal Diagram | Missing data method | | | | | | | | | | | | | | |
| | Complete case | 0.20 | 0.00 | 0.00 | -0.96 | 93.80 | 0.54 | 93.75 | 0.54 | 0.11 | 0.00 | 0.11 | 0.00 | -5.72 | 1.51 |
| | Extended TMLE | 0.20 | 0.00 | 0.00 | -0.92 | 93.60 | 0.55 | 93.85 | 0.54 | 0.11 | 0.00 | 0.11 | 0.00 | -5.27 | 1.52 |
| | Extended TMLE+MCMI | 0.22 | 0.02 | 0.00 | 10.67 | 92.00 | 0.61 | 93.00 | 0.57 | 0.10 | 0.00 | 0.10 | 0.00 | -6.16 | 1.50 |
| | MI, no int (linear) | 0.20 | 0.00 | 0.00 | -0.91 | 95.45 | 0.47 | 95.50 | 0.46 | 0.10 | 0.00 | 0.10 | 0.00 | 6.64 | 1.74 |
| T | MI,no int | 0.20 | 0.00 | 0.00 | -0.75 | 95.75 | 0.45 | 95.80 | 0.45 | 0.10 | 0.00 | 0.11 | 0.00 | 6.71 | 1.74 |
| | MI, 2-way int | 0.19 | -0.01 | 0.00 | -4.09 | 95.80 | 0.45 | 95.85 | 0.45 | 0.10 | 0.00 | 0.11 | 0.00 | 6.64 | 1.75 |
| | MI, higher int | 0.19 | -0.01 | 0.00 | -4.66 | 95.85 | 0.45 | 95.80 | 0.45 | 0.10 | 0.00 | 0.11 | 0.00 | 6.95 | 1.75 |
| | MI, CART | 0.19 | -0.01 | 0.00 | -6.27 | 95.25 | 0.48 | 95.20 | 0.48 | 0.10 | 0.00 | 0.10 | 0.00 | 2.33 | 1.67 |
| | MI, RF | 0.16 | -0.04 | 0.00 | -17.91 | 98.85 | 0.24 | 99.35 | 0.18 | 0.07 | 0.00 | 0.10 | 0.00 | 45.98 | 2.37 |
| | Complete case | 0.20 | 0.00 | 0.00 | 0.06 | 92.60 | 0.59 | 92.60 | 0.59 | 0.12 | 0.00 | 0.11 | 0.00 | -8.31 | 1.47 |
| | Extended TMLE | 0.20 | 0.00 | 0.00 | 0.38 | 92.55 | 0.59 | 92.70 | 0.58 | 0.12 | 0.00 | 0.11 | 0.00 | -8.69 | 1.46 |
| | Extended TMLE+MCMI | 0.22 | 0.02 | 0.00 | 11.18 | 92.35 | 0.59 | 93.10 | 0.57 | 0.11 | 0.00 | 0.10 | 0.00 | -8.83 | 1.46 |
| | MI, no int (linear) | 0.20 | 0.00 | 0.00 | 0.58 | 95.05 | 0.49 | 95.10 | 0.48 | 0.10 | 0.00 | 0.10 | 0.00 | 1.93 | 1.66 |
| A | MI,no int | 0.20 | 0.00 | 0.00 | -0.51 | 96.10 | 0.43 | 96.15 | 0.43 | 0.10 | 0.00 | 0.11 | 0.00 | 4.49 | 1.71 |
| | MI, 2-way int | 0.19 | -0.01 | 0.00 | -3.41 | 94.80 | 0.50 | 94.80 | 0.50 | 0.10 | 0.00 | 0.11 | 0.00 | 2.85 | 1.69 |
| | MI, higher int | 0.19 | -0.01 | 0.00 | -3.73 | 95.40 | 0.47 | 95.65 | 0.46 | 0.10 | 0.00 | 0.11 | 0.00 | 2.67 | 1.69 |
| | MI, CART | 0.19 | -0.01 | 0.00 | -5.75 | 94.75 | 0.50 | 95.25 | 0.48 | 0.10 | 0.00 | 0.10 | 0.00 | -1.05 | 1.62 |
| | MI, RF | 0.16 | -0.04 | 0.00 | -18.03 | 98.70 | 0.25 | 99.50 | 0.16 | 0.07 | 0.00 | 0.10 | 0.00 | 41.57 | 2.29 |
| | Complete case | 0.20 | 0.00 | 0.00 | 0.27 | 91.80 | 0.61 | 91.80 | 0.61 | 0.14 | 0.00 | 0.13 | 0.00 | -5.85 | 1.52 |
| | Extended TMLE | 0.20 | 0.00 | 0.00 | 0.42 | 91.55 | 0.62 | 91.55 | 0.62 | 0.14 | 0.00 | 0.13 | 0.00 | -7.16 | 1.50 |
| | Extended TMLE+MCMI | 0.22 | 0.02 | 0.00 | 8.33 | 91.15 | 0.64 | 91.55 | 0.62 | 0.13 | 0.00 | 0.12 | 0.00 | -8.67 | 1.48 |
| | MI, no int (linear) | 0.19 | -0.01 | 0.00 | -4.64 | 95.95 | 0.44 | 96.05 | 0.44 | 0.12 | 0.00 | 0.13 | 0.00 | 7.22 | 1.77 |
| B | MI,no int | 0.19 | -0.01 | 0.00 | -6.19 | 96.20 | 0.43 | 96.20 | 0.43 | 0.12 | 0.00 | 0.13 | 0.00 | 9.21 | 1.80 |
| | MI, 2-way int | 0.19 | -0.01 | 0.00 | -5.89 | 94.35 | 0.52 | 94.45 | 0.51 | 0.13 | 0.00 | 0.13 | 0.00 | 2.00 | 1.70 |
| | MI, higher int | 0.18 | -0.02 | 0.00 | -7.82 | 94.80 | 0.50 | 95.20 | 0.48 | 0.12 | 0.00 | 0.13 | 0.00 | 3.18 | 1.72 |
| | MI, CART | 0.17 | -0.03 | 0.00 | -14.95 | 94.75 | 0.50 | 95.85 | 0.45 | 0.11 | 0.00 | 0.12 | 0.00 | 8.82 | 1.81 |
| | MI, RF | 0.13 | -0.07 | 0.00 | -33.52 | 97.40 | 0.36 | 99.30 | 0.19 | 0.07 | 0.00 | 0.11 | 0.00 | 52.80 | 2.49 |
| | Complete case | 0.18 | -0.02 | 0.00 | -9.62 | 91.85 | 0.61 | 92.85 | 0.58 | 0.14 | 0.00 | 0.13 | 0.00 | -8.51 | 1.48 |
| | Extended TMLE | 0.18 | -0.02 | 0.00 | -9.69 | 92.35 | 0.59 | 92.45 | 0.59 | 0.14 | 0.00 | 0.13 | 0.00 | -9.18 | 1.47 |
| | Extended TMLE+MCMI | 0.19 | -0.01 | 0.00 | -4.39 | 92.00 | 0.61 | 92.15 | 0.60 | 0.13 | 0.00 | 0.12 | 0.00 | -9.30 | 1.46 |
| | MI, no int (linear) | 0.18 | -0.02 | 0.00 | -9.07 | 95.00 | 0.49 | 95.70 | 0.45 | 0.12 | 0.00 | 0.12 | 0.00 | 4.46 | 1.72 |
| C | MI,no int | 0.18 | -0.02 | 0.00 | -9.46 | 95.85 | 0.45 | 96.10 | 0.43 | 0.12 | 0.00 | 0.13 | 0.00 | 6.54 | 1.76 |
| | MI, 2-way int | 0.19 | -0.01 | 0.00 | -4.58 | 94.75 | 0.50 | 95.05 | 0.49 | 0.13 | 0.00 | 0.13 | 0.00 | 1.20 | 1.68 |
| | MI, higher int | 0.19 | -0.01 | 0.00 | -5.94 | 94.65 | 0.50 | 94.70 | 0.50 | 0.13 | 0.00 | 0.13 | 0.00 | 1.80 | 1.69 |
| | MI, CART | 0.17 | -0.03 | 0.00 | -17.24 | 94.40 | 0.51 | 96.05 | 0.44 | 0.11 | 0.00 | 0.12 | 0.00 | 6.30 | 1.76 |
| | MI, RF | 0.13 | -0.07 | 0.00 | -37.49 | 97.20 | 0.37 | 99.65 | 0.13 | 0.07 | 0.00 | 0.11 | 0.00 | 52.40 | 2.47 |
| | Complete case | 0.20 | 0.00 | 0.00 | 0.35 | 92.25 | 0.60 | 92.25 | 0.60 | 0.12 | 0.00 | 0.11 | 0.00 | -8.04 | 1.48 |
| | Extended TMLE | 0.20 | 0.00 | 0.00 | 0.17 | 92.50 | 0.59 | 92.50 | 0.59 | 0.12 | 0.00 | 0.11 | 0.00 | -8.04 | 1.48 |
| | Extended TMLE+MCMI | 0.22 | 0.02 | 0.00 | 12.31 | 91.95 | 0.61 | 92.75 | 0.58 | 0.11 | 0.00 | 0.10 | 0.00 | -7.81 | 1.48 |
| | MI, no int (linear) | 0.20 | 0.00 | 0.00 | 2.44 | 95.50 | 0.46 | 95.40 | 0.47 | 0.11 | 0.00 | 0.11 | 0.00 | 3.48 | 1.69 |
| D | MI,no int | 0.21 | 0.01 | 0.00 | 2.53 | 95.80 | 0.45 | 95.95 | 0.44 | 0.11 | 0.00 | 0.11 | 0.00 | 5.89 | 1.74 |
| | MI, 2-way int | 0.20 | 0.00 | 0.00 | -0.22 | 94.90 | 0.49 | 94.85 | 0.49 | 0.11 | 0.00 | 0.12 | 0.00 | 3.56 | 1.70 |
| | MI, higher int | 0.20 | 0.00 | 0.00 | 0.03 | 95.30 | 0.47 | 95.30 | 0.47 | 0.11 | 0.00 | 0.11 | 0.00 | 2.98 | 1.69 |
| | MI, CART | 0.19 | -0.01 | 0.00 | -4.30 | 94.80 | 0.50 | 94.90 | 0.49 | 0.11 | 0.00 | 0.11 | 0.00 | 1.29 | 1.66 |
| | MI, RF | 0.16 | -0.04 | 0.00 | -20.90 | 98.30 | 0.29 | 99.15 | 0.21 | 0.07 | 0.00 | 0.11 | 0.00 | 42.54 | 2.32 |
| | Complete case | 0.21 | 0.01 | 0.00 | 3.27 | 91.95 | 0.61 | 91.80 | 0.61 | 0.15 | 0.00 | 0.14 | 0.00 | -11.41 | 1.43 |
| | Extended TMLE | 0.21 | 0.01 | 0.00 | 2.99 | 91.40 | 0.63 | 91.60 | 0.62 | 0.15 | 0.00 | 0.14 | 0.00 | -11.82 | 1.43 |
| | Extended TMLE+MCMI | 0.22 | 0.02 | 0.00 | 10.40 | 91.15 | 0.64 | 91.65 | 0.62 | 0.14 | 0.00 | 0.12 | 0.00 | -11.04 | 1.44 |
| | MI, no int (linear) | 0.21 | 0.01 | 0.00 | 6.99 | 95.40 | 0.47 | 95.10 | 0.48 | 0.13 | 0.00 | 0.13 | 0.00 | 5.34 | 1.74 |
| E | MI,no int | 0.21 | 0.01 | 0.00 | 6.25 | 95.85 | 0.45 | 95.85 | 0.45 | 0.13 | 0.00 | 0.14 | 0.00 | 7.97 | 1.79 |
| | MI, 2-way int | 0.21 | 0.01 | 0.00 | 7.21 | 94.65 | 0.50 | 94.75 | 0.50 | 0.14 | 0.00 | 0.14 | 0.00 | 2.39 | 1.71 |
| | MI, higher int | 0.21 | 0.01 | 0.00 | 5.77 | 95.35 | 0.47 | 95.40 | 0.47 | 0.13 | 0.00 | 0.14 | 0.00 | 3.26 | 1.72 |
| | MI, CART | 0.18 | -0.02 | 0.00 | -7.78 | 95.30 | 0.47 | 95.65 | 0.46 | 0.12 | 0.00 | 0.13 | 0.00 | 6.45 | 1.76 |
| | MI, RF | 0.14 | -0.06 | 0.00 | -31.87 | 98.30 | 0.29 | 99.40 | 0.17 | 0.08 | 0.00 | 0.12 | 0.00 | 53.58 | 2.50 |
| | Complete case | 0.19 | -0.01 | 0.00 | -5.49 | 92.55 | 0.59 | 92.75 | 0.58 | 0.12 | 0.00 | 0.12 | 0.00 | -6.82 | 1.50 |
| | Extended TMLE | 0.19 | -0.01 | 0.00 | -5.21 | 92.35 | 0.59 | 92.60 | 0.59 | 0.12 | 0.00 | 0.12 | 0.00 | -6.55 | 1.50 |
| | Extended TMLE+MCMI | 0.21 | 0.01 | 0.00 | 7.25 | 92.55 | 0.59 | 92.70 | 0.58 | 0.11 | 0.00 | 0.10 | 0.00 | -7.95 | 1.48 |
| | MI, no int (linear) | 0.20 | 0.00 | 0.00 | -0.73 | 94.70 | 0.50 | 94.70 | 0.50 | 0.11 | 0.00 | 0.11 | 0.00 | 3.47 | 1.69 |
| F | MI,no int | 0.20 | 0.00 | 0.00 | -1.59 | 95.20 | 0.48 | 95.25 | 0.48 | 0.11 | 0.00 | 0.12 | 0.00 | 6.13 | 1.74 |
| | MI, 2-way int | 0.19 | -0.01 | 0.00 | -2.63 | 95.35 | 0.47 | 95.55 | 0.46 | 0.11 | 0.00 | 0.12 | 0.00 | 5.11 | 1.73 |
| | MI, higher int | 0.20 | 0.00 | 0.00 | -1.21 | 95.15 | 0.48 | 95.40 | 0.47 | 0.11 | 0.00 | 0.12 | 0.00 | 3.34 | 1.70 |
| | MI, CART | 0.19 | -0.01 | 0.00 | -7.09 | 95.05 | 0.49 | 95.45 | 0.47 | 0.11 | 0.00 | 0.11 | 0.00 | 4.20 | 1.71 |
| | MI, RF | 0.15 | -0.05 | 0.00 | -24.23 | 98.05 | 0.31 | 99.20 | 0.20 | 0.07 | 0.00 | 0.11 | 0.00 | 43.07 | 2.32 |
| | Complete case | 0.19 | -0.01 | 0.00 | -4.25 | 90.30 | 0.66 | 90.30 | 0.66 | 0.15 | 0.00 | 0.13 | 0.00 | -11.26 | 1.44 |
| | Extended TMLE | 0.19 | -0.01 | 0.00 | -5.09 | 90.85 | 0.64 | 90.90 | 0.64 | 0.15 | 0.00 | 0.13 | 0.00 | -10.67 | 1.44 |
| | Extended TMLE+MCMI | 0.21 | 0.01 | 0.00 | 2.65 | 91.30 | 0.63 | 91.85 | 0.61 | 0.13 | 0.00 | 0.12 | 0.00 | -9.46 | 1.46 |
| | MI, no int (linear) | 0.18 | -0.02 | 0.00 | -10.76 | 94.60 | 0.51 | 95.15 | 0.48 | 0.12 | 0.00 | 0.12 | 0.00 | 3.03 | 1.70 |
| G | MI,no int | 0.18 | -0.02 | 0.00 | -11.14 | 95.05 | 0.49 | 95.50 | 0.46 | 0.12 | 0.00 | 0.13 | 0.00 | 6.06 | 1.75 |
| | MI, 2-way int | 0.17 | -0.03 | 0.00 | -13.17 | 94.40 | 0.51 | 94.90 | 0.49 | 0.13 | 0.00 | 0.13 | 0.00 | 1.11 | 1.68 |
| | MI, higher int | 0.17 | -0.03 | 0.00 | -13.72 | 93.85 | 0.54 | 94.70 | 0.50 | 0.13 | 0.00 | 0.13 | 0.00 | 0.58 | 1.67 |
| | MI, CART | 0.16 | -0.04 | 0.00 | -19.73 | 93.95 | 0.53 | 96.05 | 0.44 | 0.11 | 0.00 | 0.12 | 0.00 | 4.15 | 1.71 |
| | MI, RF | 0.13 | -0.07 | 0.00 | -35.81 | 97.30 | 0.36 | 99.55 | 0.15 | 0.07 | 0.00 | 0.11 | 0.00 | 51.12 | 2.46 |
| | Complete case | 0.16 | -0.04 | 0.00 | -18.73 | 90.90 | 0.64 | 92.15 | 0.60 | 0.14 | 0.00 | 0.13 | 0.00 | -8.00 | 1.49 |
| | Extended TMLE | 0.16 | -0.04 | 0.00 | -19.49 | 90.40 | 0.66 | 92.05 | 0.60 | 0.14 | 0.00 | 0.13 | 0.00 | -9.13 | 1.47 |
| | Extended TMLE+MCMI | 0.17 | -0.03 | 0.00 | -14.14 | 91.15 | 0.64 | 92.30 | 0.60 | 0.13 | 0.00 | 0.11 | 0.00 | -9.55 | 1.46 |
| | MI, no int (linear) | 0.16 | -0.04 | 0.00 | -20.03 | 94.75 | 0.50 | 95.60 | 0.46 | 0.12 | 0.00 | 0.12 | 0.00 | 7.33 | 1.77 |
| H | MI,no int | 0.16 | -0.04 | 0.00 | -21.05 | 94.55 | 0.51 | 96.15 | 0.43 | 0.12 | 0.00 | 0.13 | 0.00 | 8.64 | 1.79 |
| | MI, 2-way int | 0.16 | -0.04 | 0.00 | -18.33 | 94.20 | 0.52 | 95.35 | 0.47 | 0.12 | 0.00 | 0.13 | 0.00 | 3.37 | 1.71 |
| | MI, higher int | 0.17 | -0.03 | 0.00 | -16.87 | 95.10 | 0.48 | 95.30 | 0.47 | 0.12 | 0.00 | 0.13 | 0.00 | 4.90 | 1.74 |
| | MI, CART | 0.15 | -0.05 | 0.00 | -27.23 | 93.95 | 0.53 | 96.45 | 0.41 | 0.11 | 0.00 | 0.12 | 0.00 | 8.39 | 1.79 |
| | MI, RF | 0.12 | -0.08 | 0.00 | -42.24 | 96.05 | 0.44 | 99.60 | 0.14 | 0.07 | 0.00 | 0.11 | 0.00 | 55.36 | 2.53 |
| | Complete case | 0.17 | -0.03 | 0.00 | -16.25 | 90.25 | 0.66 | 91.15 | 0.64 | 0.16 | 0.00 | 0.14 | 0.00 | -12.64 | 1.41 |
| | Extended TMLE | 0.17 | -0.03 | 0.00 | -16.43 | 89.55 | 0.68 | 90.40 | 0.66 | 0.16 | 0.00 | 0.14 | 0.00 | -13.29 | 1.40 |
| | Extended TMLE+MCMI | 0.17 | -0.03 | 0.00 | -12.81 | 90.20 | 0.66 | 91.05 | 0.64 | 0.14 | 0.00 | 0.12 | 0.00 | -12.23 | 1.42 |
| | MI, no int (linear) | 0.19 | -0.01 | 0.00 | -7.48 | 94.45 | 0.51 | 94.65 | 0.50 | 0.13 | 0.00 | 0.14 | 0.00 | 4.22 | 1.73 |
| I | MI,no int | 0.18 | -0.02 | 0.00 | -7.86 | 94.60 | 0.51 | 95.00 | 0.49 | 0.13 | 0.00 | 0.14 | 0.00 | 4.60 | 1.73 |
| | MI, 2-way int | 0.19 | -0.01 | 0.00 | -2.72 | 94.55 | 0.51 | 94.45 | 0.51 | 0.14 | 0.00 | 0.14 | 0.00 | 0.15 | 1.66 |
| | MI, higher int | 0.20 | 0.00 | 0.00 | -1.76 | 94.20 | 0.52 | 94.25 | 0.52 | 0.14 | 0.00 | 0.14 | 0.00 | 1.73 | 1.70 |
| | MI, CART | 0.16 | -0.04 | 0.00 | -20.06 | 95.30 | 0.47 | 96.05 | 0.44 | 0.12 | 0.00 | 0.13 | 0.00 | 8.75 | 1.81 |
| | MI, RF | 0.12 | -0.08 | 0.00 | -42.47 | 96.90 | 0.39 | 99.45 | 0.17 | 0.08 | 0.00 | 0.12 | 0.00 | 50.13 | 2.44 |
| | Complete case | 0.15 | -0.05 | 0.00 | -24.14 | 89.15 | 0.70 | 91.20 | 0.63 | 0.15 | 0.00 | 0.14 | 0.00 | -10.58 | 1.45 |
| | Extended TMLE | 0.15 | -0.05 | 0.00 | -25.12 | 89.00 | 0.70 | 91.00 | 0.64 | 0.15 | 0.00 | 0.14 | 0.00 | -11.66 | 1.43 |
| | Extended TMLE+MCMI | 0.16 | -0.04 | 0.00 | -20.57 | 89.15 | 0.70 | 90.45 | 0.66 | 0.14 | 0.00 | 0.12 | 0.00 | -12.80 | 1.41 |
| | MI, no int (linear) | 0.17 | -0.03 | 0.00 | -13.84 | 94.75 | 0.50 | 94.85 | 0.49 | 0.13 | 0.00 | 0.14 | 0.00 | 3.63 | 1.71 |
| J | MI,no int | 0.17 | -0.03 | 0.00 | -12.76 | 95.10 | 0.48 | 95.75 | 0.45 | 0.13 | 0.00 | 0.14 | 0.00 | 5.42 | 1.75 |
| | MI, 2-way int | 0.18 | -0.02 | 0.00 | -9.00 | 93.95 | 0.53 | 94.25 | 0.52 | 0.14 | 0.00 | 0.14 | 0.00 | -1.53 | 1.64 |
| | MI, higher int | 0.18 | -0.02 | 0.00 | -8.26 | 94.35 | 0.52 | 94.40 | 0.51 | 0.14 | 0.00 | 0.14 | 0.00 | 0.62 | 1.68 |
| | MI, CART | 0.15 | -0.05 | 0.00 | -24.68 | 94.45 | 0.51 | 96.00 | 0.44 | 0.12 | 0.00 | 0.13 | 0.00 | 8.05 | 1.80 |
| | MI, RF | 0.11 | -0.09 | 0.00 | -46.72 | 95.85 | 0.45 | 99.55 | 0.15 | 0.08 | 0.00 | 0.12 | 0.00 | 54.39 | 2.51 |

Supplementary Table 8 - Performance of missing data methods for the 11 assessed missingness mechanisms under the complex scenario 1

| | | Estimated mean difference | Absolute bias | | Relative bias (%) | Coverage (%) | | Bias eliminated coverage (%) | | Empirical SE | | Model SE | | Relative % error in model SE | |
|---|---|---|---|---|---|---|---|---|---|---|---|---|---|---|---|
| | | | Estimate | MC SE | | Estimate | MC SE | Estimate | MC SE | Estimate | MC SE | Estimate | MC SE | Estimate | MC SE |
| | Complete data | 0.20 | 0.00 | 0.00 | 2.18 | 0.93 | 0.59 | 0.92 | 0.59 | 0.08 | 0.00 | 0.08 | 0.00 | -5.89 | 1.52 |
| Causal Diagram | Missing data method | | | | | | | | | | | | | | |
| | Complete case | 0.21 | 0.01 | 0.00 | 3.36 | 91.30 | 0.63 | 91.45 | 0.63 | 0.11 | 0.00 | 0.10 | 0.00 | -11.48 | 1.43 |
| | Extended TMLE | 0.21 | 0.01 | 0.00 | 2.72 | 91.40 | 0.63 | 91.70 | 0.62 | 0.11 | 0.00 | 0.10 | 0.00 | -11.25 | 1.44 |
| | Extended TMLE+MCMI | 0.23 | 0.03 | 0.00 | 13.39 | 90.35 | 0.66 | 91.60 | 0.62 | 0.10 | 0.00 | 0.09 | 0.00 | -10.60 | 1.44 |
| | MI, no int (linear) | 0.24 | 0.04 | 0.00 | 18.02 | 95.60 | 0.46 | 96.45 | 0.41 | 0.10 | 0.00 | 0.11 | 0.00 | 12.20 | 1.84 |
| T | MI,no int | 0.23 | 0.03 | 0.00 | 15.29 | 96.05 | 0.44 | 96.65 | 0.40 | 0.10 | 0.00 | 0.11 | 0.00 | 15.51 | 1.90 |
| | MI, 2-way int | 0.19 | -0.01 | 0.00 | -3.74 | 96.90 | 0.39 | 97.00 | 0.38 | 0.10 | 0.00 | 0.11 | 0.00 | 14.92 | 1.90 |
| | MI, higher int | 0.19 | -0.01 | 0.00 | -3.64 | 96.95 | 0.38 | 97.10 | 0.38 | 0.09 | 0.00 | 0.11 | 0.00 | 17.58 | 1.94 |
| | MI, CART | 0.19 | -0.01 | 0.00 | -7.14 | 96.20 | 0.43 | 96.40 | 0.42 | 0.09 | 0.00 | 0.10 | 0.00 | 12.05 | 1.84 |
| | MI, RF | 0.15 | -0.05 | 0.00 | -23.09 | 98.40 | 0.28 | 99.65 | 0.13 | 0.06 | 0.00 | 0.10 | 0.00 | 55.88 | 2.54 |
| | Complete case | 0.21 | 0.01 | 0.00 | 5.15 | 92.10 | 0.60 | 92.45 | 0.59 | 0.11 | 0.00 | 0.10 | 0.00 | -8.64 | 1.46 |
| | Extended TMLE | 0.21 | 0.01 | 0.00 | 5.03 | 92.20 | 0.60 | 91.90 | 0.61 | 0.11 | 0.00 | 0.10 | 0.00 | -9.20 | 1.46 |
| | Extended TMLE+MCMI | 0.23 | 0.03 | 0.00 | 14.64 | 90.50 | 0.66 | 92.30 | 0.60 | 0.10 | 0.00 | 0.09 | 0.00 | -10.09 | 1.44 |
| | MI, no int (linear) | 0.23 | 0.03 | 0.00 | 14.12 | 95.95 | 0.44 | 96.85 | 0.39 | 0.10 | 0.00 | 0.11 | 0.00 | 10.31 | 1.81 |
| A | MI,no int | 0.22 | 0.02 | 0.00 | 12.43 | 96.05 | 0.44 | 97.15 | 0.37 | 0.10 | 0.00 | 0.11 | 0.00 | 12.45 | 1.85 |
| | MI, 2-way int | 0.19 | -0.01 | 0.00 | -4.69 | 96.15 | 0.43 | 96.40 | 0.42 | 0.10 | 0.00 | 0.11 | 0.00 | 11.93 | 1.84 |
| | MI, higher int | 0.19 | -0.01 | 0.00 | -2.99 | 96.75 | 0.40 | 96.85 | 0.39 | 0.10 | 0.00 | 0.11 | 0.00 | 13.90 | 1.88 |
| | MI, CART | 0.19 | -0.01 | 0.00 | -7.13 | 96.75 | 0.40 | 97.20 | 0.37 | 0.09 | 0.00 | 0.10 | 0.00 | 12.71 | 1.85 |
| | MI, RF | 0.16 | -0.04 | 0.00 | -22.07 | 98.55 | 0.27 | 99.65 | 0.13 | 0.06 | 0.00 | 0.10 | 0.00 | 53.05 | 2.49 |
| | Complete case | 0.20 | 0.00 | 0.00 | 1.85 | 90.90 | 0.64 | 90.70 | 0.65 | 0.13 | 0.00 | 0.12 | 0.00 | -10.50 | 1.44 |
| | Extended TMLE | 0.21 | 0.01 | 0.00 | 2.59 | 90.10 | 0.67 | 89.95 | 0.67 | 0.13 | 0.00 | 0.12 | 0.00 | -12.71 | 1.41 |
| | Extended TMLE+MCMI | 0.22 | 0.02 | 0.00 | 10.53 | 90.70 | 0.65 | 91.60 | 0.62 | 0.12 | 0.00 | 0.11 | 0.00 | -11.76 | 1.42 |
| | MI, no int (linear) | 0.21 | 0.01 | 0.00 | 3.95 | 96.55 | 0.41 | 96.40 | 0.42 | 0.11 | 0.00 | 0.13 | 0.00 | 13.24 | 1.88 |
| B | MI,no int | 0.21 | 0.01 | 0.00 | 2.91 | 96.90 | 0.39 | 96.95 | 0.38 | 0.11 | 0.00 | 0.13 | 0.00 | 14.37 | 1.90 |
| | MI, 2-way int | 0.18 | -0.02 | 0.00 | -9.75 | 95.70 | 0.45 | 96.05 | 0.44 | 0.12 | 0.00 | 0.13 | 0.00 | 10.86 | 1.85 |
| | MI, higher int | 0.19 | -0.01 | 0.00 | -6.79 | 96.60 | 0.41 | 96.45 | 0.41 | 0.11 | 0.00 | 0.13 | 0.00 | 13.14 | 1.89 |
| | MI, CART | 0.16 | -0.04 | 0.00 | -19.04 | 95.80 | 0.45 | 97.15 | 0.37 | 0.10 | 0.00 | 0.11 | 0.00 | 13.44 | 1.87 |
| | MI, RF | 0.12 | -0.08 | 0.00 | -38.11 | 96.85 | 0.39 | 99.85 | 0.09 | 0.07 | 0.00 | 0.11 | 0.00 | 64.71 | 2.69 |
| | Complete case | 0.19 | -0.01 | 0.00 | -6.39 | 91.30 | 0.63 | 91.75 | 0.62 | 0.13 | 0.00 | 0.12 | 0.00 | -11.61 | 1.43 |
| | Extended TMLE | 0.19 | -0.01 | 0.00 | -5.78 | 90.10 | 0.67 | 90.30 | 0.66 | 0.14 | 0.00 | 0.12 | 0.00 | -13.25 | 1.40 |
| | Extended TMLE+MCMI | 0.20 | 0.00 | 0.00 | -1.90 | 90.25 | 0.66 | 90.05 | 0.67 | 0.12 | 0.00 | 0.11 | 0.00 | -13.27 | 1.40 |
| | MI, no int (linear) | 0.20 | 0.00 | 0.00 | -0.56 | 96.35 | 0.42 | 96.40 | 0.42 | 0.11 | 0.00 | 0.13 | 0.00 | 9.56 | 1.81 |
| C | MI,no int | 0.20 | 0.00 | 0.00 | -1.83 | 96.85 | 0.39 | 96.80 | 0.39 | 0.11 | 0.00 | 0.13 | 0.00 | 11.51 | 1.84 |
| | MI, 2-way int | 0.18 | -0.02 | 0.00 | -9.51 | 95.70 | 0.45 | 96.50 | 0.41 | 0.12 | 0.00 | 0.13 | 0.00 | 9.85 | 1.83 |
| | MI, higher int | 0.19 | -0.01 | 0.00 | -6.28 | 95.80 | 0.45 | 96.15 | 0.43 | 0.12 | 0.00 | 0.13 | 0.00 | 9.48 | 1.83 |
| | MI, CART | 0.16 | -0.04 | 0.00 | -20.91 | 96.15 | 0.43 | 97.40 | 0.36 | 0.10 | 0.00 | 0.12 | 0.00 | 14.51 | 1.89 |
| | MI, RF | 0.12 | -0.08 | 0.00 | -41.35 | 96.40 | 0.42 | 99.75 | 0.11 | 0.07 | 0.00 | 0.11 | 0.00 | 59.53 | 2.61 |
| | Complete case | 0.21 | 0.01 | 0.00 | 4.15 | 92.20 | 0.60 | 92.25 | 0.60 | 0.12 | 0.00 | 0.10 | 0.00 | -9.42 | 1.45 |
| | Extended TMLE | 0.21 | 0.01 | 0.00 | 3.99 | 92.15 | 0.60 | 91.50 | 0.62 | 0.12 | 0.00 | 0.11 | 0.00 | -10.13 | 1.45 |
| | Extended TMLE+MCMI | 0.23 | 0.03 | 0.00 | 15.05 | 89.90 | 0.67 | 91.75 | 0.62 | 0.11 | 0.00 | 0.10 | 0.00 | -11.72 | 1.42 |
| | MI, no int (linear) | 0.24 | 0.04 | 0.00 | 17.86 | 94.25 | 0.52 | 95.90 | 0.44 | 0.11 | 0.00 | 0.12 | 0.00 | 8.59 | 1.78 |
| D | MI,no int | 0.23 | 0.03 | 0.00 | 15.69 | 95.45 | 0.47 | 96.55 | 0.41 | 0.10 | 0.00 | 0.12 | 0.00 | 12.00 | 1.85 |
| | MI, 2-way int | 0.20 | 0.00 | 0.00 | 0.46 | 96.60 | 0.41 | 96.40 | 0.42 | 0.11 | 0.00 | 0.12 | 0.00 | 10.48 | 1.83 |
| | MI, higher int | 0.20 | 0.00 | 0.00 | 1.38 | 96.50 | 0.41 | 96.50 | 0.41 | 0.10 | 0.00 | 0.12 | 0.00 | 13.15 | 1.87 |
| | MI, CART | 0.19 | -0.01 | 0.00 | -4.25 | 97.05 | 0.38 | 96.90 | 0.39 | 0.10 | 0.00 | 0.11 | 0.00 | 13.14 | 1.86 |
| | MI, RF | 0.15 | -0.05 | 0.00 | -23.61 | 98.80 | 0.24 | 99.90 | 0.07 | 0.07 | 0.00 | 0.11 | 0.00 | 57.59 | 2.56 |
| | Complete case | 0.21 | 0.01 | 0.00 | 3.43 | 90.80 | 0.65 | 90.50 | 0.66 | 0.14 | 0.00 | 0.12 | 0.00 | -11.90 | 1.43 |
| | Extended TMLE | 0.21 | 0.01 | 0.00 | 5.04 | 89.75 | 0.68 | 89.70 | 0.68 | 0.14 | 0.00 | 0.12 | 0.00 | -14.44 | 1.39 |
| | Extended TMLE+MCMI | 0.22 | 0.02 | 0.00 | 10.66 | 90.15 | 0.67 | 90.65 | 0.65 | 0.13 | 0.00 | 0.11 | 0.00 | -13.25 | 1.40 |
| | MI, no int (linear) | 0.23 | 0.03 | 0.00 | 13.53 | 95.45 | 0.47 | 96.40 | 0.42 | 0.12 | 0.00 | 0.13 | 0.00 | 13.17 | 1.87 |
| E | MI,no int | 0.23 | 0.03 | 0.00 | 12.59 | 96.55 | 0.41 | 96.85 | 0.39 | 0.12 | 0.00 | 0.14 | 0.00 | 14.27 | 1.90 |
| | MI, 2-way int | 0.21 | 0.01 | 0.00 | 2.71 | 96.35 | 0.42 | 96.55 | 0.41 | 0.12 | 0.00 | 0.14 | 0.00 | 10.13 | 1.84 |
| | MI, higher int | 0.21 | 0.01 | 0.00 | 3.95 | 96.00 | 0.44 | 96.10 | 0.43 | 0.12 | 0.00 | 0.13 | 0.00 | 10.78 | 1.85 |
| | MI, CART | 0.17 | -0.03 | 0.00 | -12.53 | 96.10 | 0.43 | 96.85 | 0.39 | 0.11 | 0.00 | 0.12 | 0.00 | 15.71 | 1.92 |
| | MI, RF | 0.13 | -0.07 | 0.00 | -37.37 | 98.10 | 0.31 | 99.80 | 0.10 | 0.07 | 0.00 | 0.12 | 0.00 | 62.78 | 2.65 |
| | Complete case | 0.20 | 0.00 | 0.00 | -2.01 | 91.25 | 0.63 | 91.30 | 0.63 | 0.12 | 0.00 | 0.11 | 0.00 | -11.01 | 1.43 |
| | Extended TMLE | 0.20 | 0.00 | 0.00 | -2.45 | 90.40 | 0.66 | 90.60 | 0.65 | 0.12 | 0.00 | 0.11 | 0.00 | -12.47 | 1.41 |
| | Extended TMLE+MCMI | 0.22 | 0.02 | 0.00 | 9.44 | 90.20 | 0.66 | 91.10 | 0.64 | 0.11 | 0.00 | 0.10 | 0.00 | -12.24 | 1.41 |
| | MI, no int (linear) | 0.23 | 0.03 | 0.00 | 13.16 | 95.05 | 0.49 | 95.90 | 0.44 | 0.11 | 0.00 | 0.12 | 0.00 | 8.03 | 1.77 |
| F | MI,no int | 0.22 | 0.02 | 0.00 | 11.76 | 95.20 | 0.48 | 95.80 | 0.45 | 0.11 | 0.00 | 0.12 | 0.00 | 9.49 | 1.80 |
| | MI, 2-way int | 0.20 | 0.00 | 0.00 | -1.00 | 96.20 | 0.43 | 96.45 | 0.41 | 0.11 | 0.00 | 0.12 | 0.00 | 12.12 | 1.87 |
| | MI, higher int | 0.20 | 0.00 | 0.00 | 1.66 | 97.00 | 0.38 | 96.85 | 0.39 | 0.10 | 0.00 | 0.12 | 0.00 | 13.83 | 1.88 |
| | MI, CART | 0.18 | -0.02 | 0.00 | -9.03 | 96.20 | 0.43 | 96.55 | 0.41 | 0.10 | 0.00 | 0.11 | 0.00 | 11.56 | 1.83 |
| | MI, RF | 0.15 | -0.05 | 0.00 | -26.30 | 98.10 | 0.31 | 99.40 | 0.17 | 0.07 | 0.00 | 0.11 | 0.00 | 53.13 | 2.50 |
| | Complete case | 0.20 | 0.00 | 0.00 | -1.85 | 92.10 | 0.60 | 92.10 | 0.60 | 0.13 | 0.00 | 0.12 | 0.00 | -9.36 | 1.46 |
| | Extended TMLE | 0.20 | 0.00 | 0.00 | -1.05 | 91.45 | 0.63 | 91.45 | 0.63 | 0.13 | 0.00 | 0.12 | 0.00 | -11.12 | 1.43 |
| | Extended TMLE+MCMI | 0.21 | 0.01 | 0.00 | 6.25 | 92.35 | 0.59 | 92.00 | 0.61 | 0.12 | 0.00 | 0.10 | 0.00 | -8.89 | 1.47 |
| | MI, no int (linear) | 0.20 | 0.00 | 0.00 | -1.61 | 97.15 | 0.37 | 97.25 | 0.37 | 0.11 | 0.00 | 0.13 | 0.00 | 14.29 | 1.89 |
| G | MI,no int | 0.20 | 0.00 | 0.00 | -2.30 | 96.60 | 0.41 | 96.50 | 0.41 | 0.11 | 0.00 | 0.13 | 0.00 | 15.57 | 1.91 |
| | MI, 2-way int | 0.17 | -0.03 | 0.00 | -14.05 | 95.90 | 0.44 | 96.35 | 0.42 | 0.12 | 0.00 | 0.13 | 0.00 | 10.40 | 1.83 |
| | MI, higher int | 0.18 | -0.02 | 0.00 | -11.97 | 96.50 | 0.41 | 97.55 | 0.35 | 0.11 | 0.00 | 0.12 | 0.00 | 14.16 | 1.90 |
| | MI, CART | 0.16 | -0.04 | 0.00 | -22.10 | 95.65 | 0.46 | 96.85 | 0.39 | 0.10 | 0.00 | 0.11 | 0.00 | 15.39 | 1.90 |
| | MI, RF | 0.12 | -0.08 | 0.00 | -40.21 | 96.60 | 0.41 | 99.70 | 0.12 | 0.07 | 0.00 | 0.11 | 0.00 | 63.28 | 2.66 |
| | Complete case | 0.17 | -0.03 | 0.00 | -14.83 | 90.00 | 0.67 | 90.90 | 0.64 | 0.13 | 0.00 | 0.12 | 0.00 | -10.67 | 1.44 |
| | Extended TMLE | 0.17 | -0.03 | 0.00 | -13.67 | 89.30 | 0.69 | 90.40 | 0.66 | 0.13 | 0.00 | 0.12 | 0.00 | -13.09 | 1.40 |
| | Extended TMLE+MCMI | 0.18 | -0.02 | 0.00 | -10.35 | 90.35 | 0.66 | 90.90 | 0.64 | 0.12 | 0.00 | 0.10 | 0.00 | -12.54 | 1.41 |
| | MI, no int (linear) | 0.18 | -0.02 | 0.00 | -9.55 | 96.45 | 0.41 | 96.45 | 0.41 | 0.11 | 0.00 | 0.13 | 0.00 | 11.48 | 1.84 |
| H | MI,no int | 0.18 | -0.02 | 0.00 | -9.72 | 96.65 | 0.40 | 96.45 | 0.41 | 0.11 | 0.00 | 0.13 | 0.00 | 13.38 | 1.87 |
| | MI, 2-way int | 0.17 | -0.03 | 0.00 | -16.85 | 94.85 | 0.49 | 95.95 | 0.44 | 0.12 | 0.00 | 0.13 | 0.00 | 10.55 | 1.84 |
| | MI, higher int | 0.17 | -0.03 | 0.00 | -13.74 | 95.40 | 0.47 | 95.85 | 0.45 | 0.11 | 0.00 | 0.12 | 0.00 | 9.22 | 1.82 |
| | MI, CART | 0.15 | -0.05 | 0.00 | -27.16 | 95.00 | 0.49 | 97.30 | 0.36 | 0.10 | 0.00 | 0.12 | 0.00 | 14.77 | 1.90 |
| | MI, RF | 0.11 | -0.09 | 0.00 | -45.62 | 95.50 | 0.46 | 99.80 | 0.10 | 0.07 | 0.00 | 0.11 | 0.00 | 61.42 | 2.62 |
| | Complete case | 0.18 | -0.02 | 0.00 | -11.89 | 90.70 | 0.65 | 91.40 | 0.63 | 0.14 | 0.00 | 0.12 | 0.00 | -13.14 | 1.40 |
| | Extended TMLE | 0.18 | -0.02 | 0.00 | -11.07 | 89.35 | 0.69 | 89.55 | 0.68 | 0.14 | 0.00 | 0.12 | 0.00 | -16.06 | 1.36 |
| | Extended TMLE+MCMI | 0.18 | -0.02 | 0.00 | -9.17 | 89.45 | 0.69 | 89.85 | 0.68 | 0.13 | 0.00 | 0.11 | 0.00 | -15.58 | 1.36 |
| | MI, no int (linear) | 0.20 | 0.00 | 0.00 | 2.45 | 96.20 | 0.43 | 96.05 | 0.44 | 0.12 | 0.00 | 0.14 | 0.00 | 10.97 | 1.83 |
| I | MI,no int | 0.20 | 0.00 | 0.00 | 2.27 | 96.90 | 0.39 | 96.65 | 0.40 | 0.12 | 0.00 | 0.14 | 0.00 | 13.90 | 1.91 |
| | MI, 2-way int | 0.20 | 0.00 | 0.00 | -1.98 | 95.35 | 0.47 | 95.20 | 0.48 | 0.12 | 0.00 | 0.14 | 0.00 | 10.69 | 1.86 |
| | MI, higher int | 0.20 | 0.00 | 0.00 | 0.20 | 96.45 | 0.41 | 96.45 | 0.41 | 0.12 | 0.00 | 0.13 | 0.00 | 9.61 | 1.83 |
| | MI, CART | 0.16 | -0.04 | 0.00 | -20.32 | 95.35 | 0.47 | 97.05 | 0.38 | 0.11 | 0.00 | 0.12 | 0.00 | 14.39 | 1.89 |
| | MI, RF | 0.11 | -0.09 | 0.00 | -45.07 | 96.35 | 0.42 | 99.75 | 0.11 | 0.07 | 0.00 | 0.12 | 0.00 | 62.56 | 2.66 |
| | Complete case | 0.16 | -0.04 | 0.00 | -20.90 | 89.15 | 0.70 | 91.35 | 0.63 | 0.14 | 0.00 | 0.12 | 0.00 | -12.73 | 1.41 |
| | Extended TMLE | 0.16 | -0.04 | 0.00 | -20.32 | 88.40 | 0.72 | 89.85 | 0.68 | 0.14 | 0.00 | 0.12 | 0.00 | -14.56 | 1.38 |
| | Extended TMLE+MCMI | 0.16 | -0.04 | 0.00 | -17.99 | 88.15 | 0.72 | 90.15 | 0.67 | 0.13 | 0.00 | 0.11 | 0.00 | -14.11 | 1.38 |
| | MI, no int (linear) | 0.19 | -0.01 | 0.00 | -5.13 | 96.80 | 0.39 | 96.85 | 0.39 | 0.12 | 0.00 | 0.14 | 0.00 | 14.27 | 1.90 |
| J | MI,no int | 0.19 | -0.01 | 0.00 | -6.78 | 96.70 | 0.40 | 96.80 | 0.39 | 0.12 | 0.00 | 0.14 | 0.00 | 14.54 | 1.90 |
| | MI, 2-way int | 0.18 | -0.02 | 0.00 | -10.10 | 95.55 | 0.46 | 96.10 | 0.43 | 0.12 | 0.00 | 0.14 | 0.00 | 9.92 | 1.83 |
| | MI, higher int | 0.19 | -0.01 | 0.00 | -5.92 | 97.10 | 0.38 | 97.50 | 0.35 | 0.12 | 0.00 | 0.13 | 0.00 | 12.28 | 1.87 |
| | MI, CART | 0.15 | -0.05 | 0.00 | -27.37 | 95.15 | 0.48 | 97.50 | 0.35 | 0.11 | 0.00 | 0.12 | 0.00 | 16.28 | 1.92 |
| | MI, RF | 0.10 | -0.10 | 0.00 | -49.35 | 96.30 | 0.42 | 99.80 | 0.10 | 0.07 | 0.00 | 0.12 | 0.00 | 64.96 | 2.70 |

Supplementary Table 9 - Performance of missing data methods for the 11 assessed missingness mechanisms under the complex scenario 2

| | | Estimated mean difference | Absolute bias | | Relative bias (%) | Coverage (%) | | Bias eliminated coverage (%) | | Empirical SE | | Model SE | | Relative % error in model SE | |
|---|---|---|---|---|---|---|---|---|---|---|---|---|---|---|---|
| | | | Estimate | MC SE | | Estimate | MC SE | Estimate | MC SE | Estimate | MC SE | Estimate | MC SE | Estimate | MC SE |
| | Complete data | 0.20 | 0.00 | 0.00 | 0.33 | 0.87 | 0.76 | 0.87 | 0.75 | 0.09 | 0.00 | 0.07 | 0.00 | -20.94 | 1.29 |
| Causal Diagram | Missing data method | | | | | | | | | | | | | | |
| | Complete case | 0.21 | 0.01 | 0.00 | 3.51 | 85.85 | 0.78 | 86.55 | 0.76 | 0.13 | 0.00 | 0.10 | 0.00 | -24.47 | 1.23 |
| | Extended TMLE | 0.21 | 0.01 | 0.00 | 3.68 | 85.40 | 0.79 | 85.20 | 0.79 | 0.13 | 0.00 | 0.10 | 0.00 | -24.61 | 1.23 |
| | Extended TMLE+MCMI | 0.23 | 0.03 | 0.00 | 15.53 | 83.70 | 0.83 | 85.70 | 0.78 | 0.11 | 0.00 | 0.09 | 0.00 | -23.81 | 1.23 |
| | MI, no int (linear) | 0.32 | 0.12 | 0.00 | 59.49 | 90.70 | 0.65 | 97.75 | 0.33 | 0.11 | 0.00 | 0.13 | 0.00 | 25.32 | 2.08 |
| T | MI,no int | 0.31 | 0.11 | 0.00 | 55.29 | 91.90 | 0.61 | 97.50 | 0.35 | 0.11 | 0.00 | 0.14 | 0.00 | 27.06 | 2.12 |
| | MI, 2-way int | 0.21 | 0.01 | 0.00 | 5.91 | 98.40 | 0.28 | 97.95 | 0.32 | 0.10 | 0.00 | 0.13 | 0.00 | 35.17 | 2.26 |
| | MI, higher int | 0.21 | 0.01 | 0.00 | 6.51 | 98.60 | 0.26 | 98.60 | 0.26 | 0.10 | 0.00 | 0.13 | 0.00 | 34.77 | 2.25 |
| | MI, CART | 0.20 | 0.00 | 0.00 | -0.26 | 97.80 | 0.33 | 97.80 | 0.33 | 0.10 | 0.00 | 0.12 | 0.00 | 25.12 | 2.08 |
| | MI, RF | 0.16 | -0.04 | 0.00 | -20.74 | 99.40 | 0.17 | 99.90 | 0.07 | 0.07 | 0.00 | 0.12 | 0.00 | 78.06 | 2.94 |
| | Complete case | 0.20 | 0.00 | 0.00 | 2.06 | 90.50 | 0.66 | 90.20 | 0.66 | 0.11 | 0.00 | 0.09 | 0.00 | -13.71 | 1.39 |
| | Extended TMLE | 0.21 | 0.01 | 0.00 | 2.59 | 89.85 | 0.68 | 89.60 | 0.68 | 0.11 | 0.00 | 0.09 | 0.00 | -14.65 | 1.38 |
| | Extended TMLE+MCMI | 0.23 | 0.03 | 0.00 | 13.19 | 88.50 | 0.71 | 89.60 | 0.68 | 0.10 | 0.00 | 0.09 | 0.00 | -16.78 | 1.34 |
| | MI, no int (linear) | 0.29 | 0.09 | 0.00 | 43.59 | 93.75 | 0.54 | 97.15 | 0.37 | 0.11 | 0.00 | 0.13 | 0.00 | 20.44 | 2.00 |
| A | MI,no int | 0.28 | 0.08 | 0.00 | 41.77 | 94.50 | 0.51 | 98.05 | 0.31 | 0.11 | 0.00 | 0.14 | 0.00 | 25.05 | 2.08 |
| | MI, 2-way int | 0.22 | 0.02 | 0.00 | 8.12 | 98.60 | 0.26 | 98.65 | 0.26 | 0.10 | 0.00 | 0.14 | 0.00 | 37.56 | 2.30 |
| | MI, higher int | 0.22 | 0.02 | 0.00 | 11.31 | 98.40 | 0.28 | 98.70 | 0.25 | 0.10 | 0.00 | 0.13 | 0.00 | 35.82 | 2.26 |
| | MI, CART | 0.20 | 0.00 | 0.00 | -2.29 | 98.65 | 0.26 | 98.65 | 0.26 | 0.09 | 0.00 | 0.12 | 0.00 | 26.99 | 2.12 |
| | MI, RF | 0.16 | -0.04 | 0.00 | -20.18 | 99.15 | 0.21 | 99.60 | 0.14 | 0.07 | 0.00 | 0.11 | 0.00 | 67.82 | 2.75 |
| | Complete case | 0.20 | 0.00 | 0.00 | -0.34 | 89.70 | 0.68 | 89.60 | 0.68 | 0.13 | 0.00 | 0.11 | 0.00 | -15.37 | 1.37 |
| | Extended TMLE | 0.20 | 0.00 | 0.00 | 1.68 | 88.25 | 0.72 | 88.15 | 0.72 | 0.13 | 0.00 | 0.11 | 0.00 | -18.14 | 1.33 |
| | Extended TMLE+MCMI | 0.22 | 0.02 | 0.00 | 10.55 | 87.75 | 0.73 | 88.75 | 0.71 | 0.12 | 0.00 | 0.10 | 0.00 | -18.49 | 1.32 |
| | MI, no int (linear) | 0.26 | 0.06 | 0.00 | 27.76 | 96.40 | 0.42 | 97.25 | 0.37 | 0.12 | 0.00 | 0.15 | 0.00 | 21.11 | 2.01 |
| B | MI,no int | 0.26 | 0.06 | 0.00 | 29.30 | 96.10 | 0.43 | 97.10 | 0.38 | 0.13 | 0.00 | 0.15 | 0.00 | 21.31 | 2.03 |
| | MI, 2-way int | 0.20 | 0.00 | 0.00 | 1.35 | 98.00 | 0.31 | 97.95 | 0.32 | 0.12 | 0.00 | 0.15 | 0.00 | 28.04 | 2.14 |
| | MI, higher int | 0.21 | 0.01 | 0.00 | 5.83 | 97.95 | 0.32 | 98.00 | 0.31 | 0.11 | 0.00 | 0.14 | 0.00 | 28.10 | 2.14 |
| | MI, CART | 0.17 | -0.03 | 0.00 | -12.96 | 97.50 | 0.35 | 98.00 | 0.31 | 0.10 | 0.00 | 0.13 | 0.00 | 23.40 | 2.05 |
| | MI, RF | 0.12 | -0.08 | 0.00 | -39.14 | 97.75 | 0.33 | 99.85 | 0.09 | 0.07 | 0.00 | 0.12 | 0.00 | 73.54 | 2.84 |
| | Complete case | 0.18 | -0.02 | 0.00 | -11.52 | 89.65 | 0.68 | 89.75 | 0.68 | 0.13 | 0.00 | 0.11 | 0.00 | -14.89 | 1.38 |
| | Extended TMLE | 0.18 | -0.02 | 0.00 | -9.96 | 88.15 | 0.72 | 87.80 | 0.73 | 0.13 | 0.00 | 0.11 | 0.00 | -18.17 | 1.33 |
| | Extended TMLE+MCMI | 0.19 | -0.01 | 0.00 | -4.30 | 88.40 | 0.72 | 88.25 | 0.72 | 0.12 | 0.00 | 0.10 | 0.00 | -18.32 | 1.32 |
| | MI, no int (linear) | 0.24 | 0.04 | 0.00 | 22.12 | 96.45 | 0.41 | 97.40 | 0.36 | 0.12 | 0.00 | 0.15 | 0.00 | 20.25 | 1.99 |
| C | MI,no int | 0.25 | 0.05 | 0.00 | 23.63 | 97.55 | 0.35 | 97.65 | 0.34 | 0.12 | 0.00 | 0.15 | 0.00 | 23.74 | 2.07 |
| | MI, 2-way int | 0.21 | 0.01 | 0.00 | 2.86 | 97.95 | 0.32 | 98.05 | 0.31 | 0.12 | 0.00 | 0.15 | 0.00 | 27.88 | 2.14 |
| | MI, higher int | 0.22 | 0.02 | 0.00 | 9.06 | 97.85 | 0.32 | 98.10 | 0.31 | 0.11 | 0.00 | 0.14 | 0.00 | 28.89 | 2.16 |
| | MI, CART | 0.17 | -0.03 | 0.00 | -15.27 | 96.80 | 0.39 | 97.35 | 0.36 | 0.10 | 0.00 | 0.13 | 0.00 | 24.03 | 2.06 |
| | MI, RF | 0.12 | -0.08 | 0.00 | -40.75 | 98.00 | 0.31 | 99.80 | 0.10 | 0.07 | 0.00 | 0.12 | 0.00 | 76.85 | 2.91 |
| | Complete case | 0.20 | 0.00 | 0.00 | 1.33 | 90.20 | 0.66 | 90.15 | 0.67 | 0.12 | 0.00 | 0.10 | 0.00 | -16.25 | 1.35 |
| | Extended TMLE | 0.20 | 0.00 | 0.00 | 1.81 | 89.10 | 0.70 | 89.00 | 0.70 | 0.12 | 0.00 | 0.10 | 0.00 | -17.88 | 1.33 |
| | Extended TMLE+MCMI | 0.23 | 0.03 | 0.00 | 13.57 | 88.25 | 0.72 | 89.30 | 0.69 | 0.11 | 0.00 | 0.09 | 0.00 | -17.87 | 1.32 |
| | MI, no int (linear) | 0.29 | 0.09 | 0.00 | 44.82 | 92.90 | 0.57 | 97.40 | 0.36 | 0.12 | 0.00 | 0.14 | 0.00 | 18.38 | 1.96 |
| D | MI,no int | 0.28 | 0.08 | 0.00 | 41.24 | 94.50 | 0.51 | 97.65 | 0.34 | 0.11 | 0.00 | 0.14 | 0.00 | 23.33 | 2.05 |
| | MI, 2-way int | 0.23 | 0.03 | 0.00 | 13.11 | 97.65 | 0.34 | 98.05 | 0.31 | 0.11 | 0.00 | 0.14 | 0.00 | 31.40 | 2.20 |
| | MI, higher int | 0.23 | 0.03 | 0.00 | 16.60 | 98.10 | 0.31 | 98.20 | 0.30 | 0.10 | 0.00 | 0.14 | 0.00 | 31.61 | 2.20 |
| | MI, CART | 0.20 | 0.00 | 0.00 | -0.09 | 98.25 | 0.29 | 98.25 | 0.29 | 0.10 | 0.00 | 0.12 | 0.00 | 25.02 | 2.07 |
| | MI, RF | 0.15 | -0.05 | 0.00 | -24.21 | 99.35 | 0.18 | 99.80 | 0.10 | 0.07 | 0.00 | 0.12 | 0.00 | 72.10 | 2.83 |
| | Complete case | 0.19 | -0.01 | 0.00 | -3.51 | 89.75 | 0.68 | 89.80 | 0.68 | 0.13 | 0.00 | 0.11 | 0.00 | -15.11 | 1.37 |
| | Extended TMLE | 0.20 | 0.00 | 0.00 | -1.92 | 88.25 | 0.72 | 88.05 | 0.73 | 0.14 | 0.00 | 0.11 | 0.00 | -19.22 | 1.31 |
| | Extended TMLE+MCMI | 0.21 | 0.01 | 0.00 | 6.25 | 88.15 | 0.72 | 88.25 | 0.72 | 0.13 | 0.00 | 0.10 | 0.00 | -19.62 | 1.30 |
| | MI, no int (linear) | 0.27 | 0.07 | 0.00 | 33.01 | 96.50 | 0.41 | 97.15 | 0.37 | 0.13 | 0.00 | 0.15 | 0.00 | 20.37 | 2.01 |
| E | MI,no int | 0.27 | 0.07 | 0.00 | 32.60 | 95.85 | 0.45 | 97.25 | 0.37 | 0.13 | 0.00 | 0.16 | 0.00 | 22.28 | 2.05 |
| | MI, 2-way int | 0.21 | 0.01 | 0.00 | 6.93 | 97.65 | 0.34 | 98.05 | 0.31 | 0.12 | 0.00 | 0.15 | 0.00 | 29.97 | 2.18 |
| | MI, higher int | 0.22 | 0.02 | 0.00 | 12.08 | 98.20 | 0.30 | 98.20 | 0.30 | 0.11 | 0.00 | 0.15 | 0.00 | 32.82 | 2.23 |
| | MI, CART | 0.18 | -0.02 | 0.00 | -11.32 | 97.30 | 0.36 | 97.75 | 0.33 | 0.10 | 0.00 | 0.13 | 0.00 | 27.81 | 2.13 |
| | MI, RF | 0.12 | -0.08 | 0.00 | -41.37 | 98.35 | 0.28 | 99.85 | 0.09 | 0.07 | 0.00 | 0.13 | 0.00 | 80.43 | 2.98 |
| | Complete case | 0.19 | -0.01 | 0.00 | -5.16 | 90.50 | 0.66 | 90.70 | 0.65 | 0.12 | 0.00 | 0.10 | 0.00 | -14.86 | 1.37 |
| | Extended TMLE | 0.19 | -0.01 | 0.00 | -4.94 | 89.30 | 0.69 | 89.80 | 0.68 | 0.12 | 0.00 | 0.10 | 0.00 | -17.00 | 1.34 |
| | Extended TMLE+MCMI | 0.22 | 0.02 | 0.00 | 8.67 | 88.05 | 0.73 | 88.35 | 0.72 | 0.11 | 0.00 | 0.09 | 0.00 | -17.80 | 1.32 |
| | MI, no int (linear) | 0.28 | 0.08 | 0.00 | 41.97 | 94.45 | 0.51 | 97.15 | 0.37 | 0.12 | 0.00 | 0.14 | 0.00 | 18.59 | 1.96 |
| F | MI,no int | 0.28 | 0.08 | 0.00 | 39.10 | 94.85 | 0.49 | 97.20 | 0.37 | 0.11 | 0.00 | 0.14 | 0.00 | 25.54 | 2.10 |
| | MI, 2-way int | 0.23 | 0.03 | 0.00 | 13.99 | 98.05 | 0.31 | 98.15 | 0.30 | 0.11 | 0.00 | 0.14 | 0.00 | 30.09 | 2.16 |
| | MI, higher int | 0.24 | 0.04 | 0.00 | 18.81 | 97.35 | 0.36 | 97.95 | 0.32 | 0.10 | 0.00 | 0.14 | 0.00 | 31.23 | 2.19 |
| | MI, CART | 0.20 | 0.00 | 0.00 | -0.43 | 97.55 | 0.35 | 97.55 | 0.35 | 0.10 | 0.00 | 0.12 | 0.00 | 24.99 | 2.07 |
| | MI, RF | 0.15 | -0.05 | 0.00 | -23.55 | 99.55 | 0.15 | 99.85 | 0.09 | 0.07 | 0.00 | 0.12 | 0.00 | 72.62 | 2.84 |
| | Complete case | 0.18 | -0.02 | 0.00 | -9.28 | 89.85 | 0.68 | 91.10 | 0.64 | 0.13 | 0.00 | 0.11 | 0.00 | -14.62 | 1.38 |
| | Extended TMLE | 0.18 | -0.02 | 0.00 | -7.92 | 88.80 | 0.71 | 89.30 | 0.69 | 0.13 | 0.00 | 0.11 | 0.00 | -18.04 | 1.33 |
| | Extended TMLE+MCMI | 0.20 | 0.00 | 0.00 | 1.77 | 88.40 | 0.72 | 88.30 | 0.72 | 0.12 | 0.00 | 0.10 | 0.00 | -17.01 | 1.34 |
| | MI, no int (linear) | 0.24 | 0.04 | 0.00 | 19.64 | 96.70 | 0.40 | 97.70 | 0.34 | 0.12 | 0.00 | 0.15 | 0.00 | 21.75 | 2.02 |
| G | MI,no int | 0.24 | 0.04 | 0.00 | 20.30 | 97.15 | 0.37 | 97.70 | 0.34 | 0.12 | 0.00 | 0.15 | 0.00 | 23.80 | 2.07 |
| | MI, 2-way int | 0.19 | -0.01 | 0.00 | -7.17 | 98.10 | 0.31 | 98.05 | 0.31 | 0.11 | 0.00 | 0.15 | 0.00 | 31.72 | 2.21 |
| | MI, higher int | 0.20 | 0.00 | 0.00 | -1.70 | 97.70 | 0.34 | 97.65 | 0.34 | 0.11 | 0.00 | 0.14 | 0.00 | 31.20 | 2.21 |
| | MI, CART | 0.16 | -0.04 | 0.00 | -18.82 | 96.65 | 0.40 | 98.00 | 0.31 | 0.10 | 0.00 | 0.13 | 0.00 | 28.36 | 2.14 |
| | MI, RF | 0.11 | -0.09 | 0.00 | -42.84 | 97.20 | 0.37 | 99.90 | 0.07 | 0.07 | 0.00 | 0.12 | 0.00 | 78.17 | 2.92 |
| | Complete case | 0.16 | -0.04 | 0.00 | -17.51 | 89.80 | 0.68 | 90.20 | 0.66 | 0.13 | 0.00 | 0.11 | 0.00 | -14.01 | 1.39 |
| | Extended TMLE | 0.17 | -0.03 | 0.00 | -16.62 | 88.55 | 0.71 | 89.10 | 0.70 | 0.13 | 0.00 | 0.11 | 0.00 | -17.71 | 1.33 |
| | Extended TMLE+MCMI | 0.18 | -0.02 | 0.00 | -9.88 | 89.35 | 0.69 | 90.00 | 0.67 | 0.12 | 0.00 | 0.10 | 0.00 | -16.33 | 1.35 |
| | MI, no int (linear) | 0.23 | 0.03 | 0.00 | 14.25 | 97.10 | 0.38 | 97.10 | 0.38 | 0.12 | 0.00 | 0.15 | 0.00 | 23.34 | 2.06 |
| H | MI,no int | 0.23 | 0.03 | 0.00 | 16.38 | 97.70 | 0.34 | 97.80 | 0.33 | 0.12 | 0.00 | 0.15 | 0.00 | 25.05 | 2.09 |
| | MI, 2-way int | 0.19 | -0.01 | 0.00 | -2.98 | 98.35 | 0.28 | 98.40 | 0.28 | 0.11 | 0.00 | 0.15 | 0.00 | 29.30 | 2.16 |
| | MI, higher int | 0.21 | 0.01 | 0.00 | 3.13 | 97.70 | 0.34 | 97.70 | 0.34 | 0.11 | 0.00 | 0.14 | 0.00 | 31.26 | 2.20 |
| | MI, CART | 0.16 | -0.04 | 0.00 | -20.48 | 97.15 | 0.37 | 98.05 | 0.31 | 0.10 | 0.00 | 0.12 | 0.00 | 26.72 | 2.11 |
| | MI, RF | 0.11 | -0.09 | 0.00 | -44.17 | 97.15 | 0.37 | 99.85 | 0.09 | 0.07 | 0.00 | 0.12 | 0.00 | 77.77 | 2.91 |
| | Complete case | 0.16 | -0.04 | 0.00 | -18.08 | 88.10 | 0.72 | 89.15 | 0.70 | 0.13 | 0.00 | 0.11 | 0.00 | -15.61 | 1.37 |
| | Extended TMLE | 0.17 | -0.03 | 0.00 | -17.40 | 86.65 | 0.76 | 87.90 | 0.73 | 0.14 | 0.00 | 0.11 | 0.00 | -18.99 | 1.31 |
| | Extended TMLE+MCMI | 0.17 | -0.03 | 0.00 | -13.79 | 87.40 | 0.74 | 87.85 | 0.73 | 0.12 | 0.00 | 0.10 | 0.00 | -18.78 | 1.31 |
| | MI, no int (linear) | 0.24 | 0.04 | 0.00 | 22.28 | 96.70 | 0.40 | 97.40 | 0.36 | 0.13 | 0.00 | 0.16 | 0.00 | 21.72 | 2.03 |
| I | MI,no int | 0.25 | 0.05 | 0.00 | 23.23 | 96.85 | 0.39 | 97.70 | 0.34 | 0.13 | 0.00 | 0.16 | 0.00 | 22.45 | 2.05 |
| | MI, 2-way int | 0.21 | 0.01 | 0.00 | 4.35 | 97.85 | 0.32 | 97.85 | 0.32 | 0.12 | 0.00 | 0.15 | 0.00 | 26.99 | 2.13 |
| | MI, higher int | 0.22 | 0.02 | 0.00 | 11.32 | 97.80 | 0.33 | 97.80 | 0.33 | 0.12 | 0.00 | 0.15 | 0.00 | 27.35 | 2.13 |
| | MI, CART | 0.17 | -0.03 | 0.00 | -16.78 | 97.15 | 0.37 | 97.75 | 0.33 | 0.11 | 0.00 | 0.13 | 0.00 | 24.87 | 2.08 |
| | MI, RF | 0.11 | -0.09 | 0.00 | -46.49 | 97.90 | 0.32 | 99.80 | 0.10 | 0.07 | 0.00 | 0.13 | 0.00 | 83.13 | 3.01 |
| | Complete case | 0.15 | -0.05 | 0.00 | -23.42 | 86.70 | 0.76 | 89.00 | 0.70 | 0.14 | 0.00 | 0.11 | 0.00 | -16.41 | 1.35 |
| | Extended TMLE | 0.15 | -0.05 | 0.00 | -22.61 | 86.15 | 0.77 | 88.15 | 0.72 | 0.14 | 0.00 | 0.11 | 0.00 | -19.31 | 1.31 |
| | Extended TMLE+MCMI | 0.16 | -0.04 | 0.00 | -17.81 | 86.10 | 0.77 | 88.35 | 0.72 | 0.12 | 0.00 | 0.10 | 0.00 | -18.59 | 1.32 |
| | MI, no int (linear) | 0.23 | 0.03 | 0.00 | 17.04 | 96.65 | 0.40 | 96.90 | 0.39 | 0.13 | 0.00 | 0.16 | 0.00 | 19.18 | 1.99 |
| J | MI,no int | 0.24 | 0.04 | 0.00 | 18.56 | 96.55 | 0.41 | 97.10 | 0.38 | 0.13 | 0.00 | 0.16 | 0.00 | 21.48 | 2.05 |
| | MI, 2-way int | 0.20 | 0.00 | 0.00 | -0.16 | 98.30 | 0.29 | 98.30 | 0.29 | 0.12 | 0.00 | 0.15 | 0.00 | 26.74 | 2.13 |
| | MI, higher int | 0.21 | 0.01 | 0.00 | 7.09 | 97.40 | 0.36 | 97.40 | 0.36 | 0.12 | 0.00 | 0.15 | 0.00 | 25.91 | 2.12 |
| | MI, CART | 0.16 | -0.04 | 0.00 | -20.79 | 96.70 | 0.40 | 98.05 | 0.31 | 0.10 | 0.00 | 0.13 | 0.00 | 29.13 | 2.15 |
| | MI, RF | 0.10 | -0.10 | 0.00 | -49.99 | 97.15 | 0.37 | 99.65 | 0.13 | 0.07 | 0.00 | 0.13 | 0.00 | 79.63 | 2.95 |